\documentclass[aps,pre,notitlepage,floatfix,groupedaddress,10pt]{revtex4-2}

\usepackage{amsmath}
\usepackage{amsfonts}
\usepackage{amssymb}
\usepackage{graphicx}
\usepackage{color}
\usepackage{hyperref}
\usepackage{bm}
\usepackage[normalem]{ulem}
\usepackage{overpic}
\usepackage[capitalize]{cleveref}

\definecolor{darkblue}{rgb}{0.00,0.00,0.55}
\definecolor{black}{rgb}{0.00,0.00,0.00}
\hypersetup{
    linkcolor = darkblue,
    anchorcolor = darkblue,
    citecolor = darkblue,
    filecolor = darkblue,
    urlcolor = darkblue,
    colorlinks  = true,
}
\definecolor{brightcerulean}{rgb}{0.11, 0.67, 0.84}
\DeclareMathAlphabet\mathbfcal{OMS}{cmsy}{b}{n}

\newcounter{fig}

\newcommand{\avg}[1]{\langle{#1}\rangle}

\newcommand{\pd}{\partial}

\newcommand{\vf}[1]{{\bf{#1}}}
\newcommand{\figwidth}{0.9\textwidth}

\begin{document}

\title{Bifurcation analysis of two-dimensional Rayleigh--B\'enard convection using deflation}

\author{N.~Boull\'e}
\email[Email: ]{nicolas.boulle@maths.ox.ac.uk}
\affiliation{Mathematical Institute, University of Oxford, Woodstock Road, Oxford OX2 6GG, UK}
\author{V.~Dallas}
\email[Email: ]{vassilios.dallas@maths.ox.ac.uk}
\affiliation{Mathematical Institute, University of Oxford, Woodstock Road, Oxford OX2 6GG, UK}
\author{P.~E.~Farrell}
\email[Email: ]{patrick.farrell@maths.ox.ac.uk}
\affiliation{Mathematical Institute, University of Oxford, Woodstock Road, Oxford OX2 6GG, UK}

\date{\today}

\begin{abstract}
We perform a bifurcation analysis of the steady states of Rayleigh--B\'enard convection with no-slip boundary conditions in two dimensions using a numerical method called deflated continuation. By combining this method with an initialisation strategy based on the eigenmodes of the conducting state, we are able to discover multiple solutions to this non-linear problem, including disconnected branches of the bifurcation diagram, without the need for any prior knowledge of the solutions. One of the disconnected branches we find contains an S-shaped curve with hysteresis, which is the origin of a flow pattern that may be related to the dynamics of flow reversals in the turbulent regime. Linear stability analysis is also performed to analyse the steady and unsteady regimes of the solutions in the parameter space and to characterise the type of instabilities.
\end{abstract}

\maketitle

\section{Introduction}

This paper considers the bifurcations of the two-dimensional Rayleigh--B\'enard convection, which supports a profusion of states. There have been numerous investigations trying to visualise solutions and characterise the steady states of Rayleigh--B\'enard convection, as well as computing its bifurcation diagrams with respect to $Ra$, the Rayleigh number, in various settings and for different geometries~\cite{crosshohenberg93,bodenschatzetal00}. Ouertatani et al.~\cite{ouertatani2008numerical} performed numerical simulations of Rayleigh--B\'enard convection in a square cavity using the finite volume method and computed fluid flow profiles and temperature patterns at $Ra=10^4,10^5,10^6$. Numerical studies have been conducted to understand the existence of bifurcating solutions to this problem in a two-dimensional domain with periodic boundary conditions in the horizontal direction~\cite{zienicke1998bifurcations,paul2012bifurcation}. Mishra et al.~\cite{mishra2010patterns} analysed the effect of low Prandtl number on the bifurcation structures, while Peterson~\cite{Peterson2008} used arclength continuation~\cite{keller1977numerical} to study the evolution of cell solutions with respect to the aspect ratio of the domain. Bifurcation structures of Rayleigh--B\'enard convection have been extensively studied in cylindrical geometry by Ma et al.~\cite{ma2006multiplicity} as well as by Boro{\'n}ska and Tuckerman~\cite{boronska2010extreme,boronska2010extreme2}. In particular, references~\cite{boronska2010extreme,boronska2010extreme2} adapted a time-dependent code to perform branch continuation and then analysed the linear stability of the computed solutions using Arnoldi iterations~\cite{arnoldi1951principle}. Finally, Puigjaner et al.~\cite{puigjaner2004stability,puigjaner2006bifurcation} computed bifurcation diagrams for Rayleigh--B\'enard convection in a cubical cavity over different intervals of $Ra < 1.5 \times 10^5$.

A common way to reconstruct bifurcation diagrams is to use arclength continuation and branch switching techniques~\cite{doedel1981auto,keller1977numerical,uecker2014pde2path} to continue known solutions in a given parameter. These solutions can be computed by applying Newton's method to suitably chosen initial guesses or by using time-evolution algorithms. A standard numerical technique in fluid dynamics for finding initial states for the continuation algorithm is to perform time-dependent simulation of initial solutions until a steady state is reached~\cite{dijkstra2014numerical,tuckerman2000bifurcation,mamun1995asymmetry}. These numerical techniques have been widely used and are extremely successful at computing bifurcation diagrams of nonlinear PDEs.

In this paper, we employ a recently developed algorithm called deflation for computing multiple solutions to nonlinear partial differential equations~\cite{farrell2015deflation,farrell2016computation}. This method has been successfully applied to compute bifurcations to a wide range of physical problems such as the deformation of a hyperelastic beam~\cite{farrell2016computation}, Carrier's problem~\cite{chapman2017analysis}, cholesteric liquid crystals~\cite{emerson2018computing}, and the nonlinear Schr\"odinger equation in two and three dimensions~\cite{charalampidis2018computing,charalampidis2019bifurcation,boulle2020deflation}. We emphasise that deflation offers some advantages that could be combined with the standard bifurcation analysis tools. The main strength of deflation is that it detects of disconnected branches, in addition to those connected to known branches. An additional advantage is that it does not require the solution of augmented problems to find new solutions. Using perturbed solutions to the linearised equations as initial conditions, we are able to discover numerous steady states of Rayleigh--B\'enard convection. These solutions can be obtained without prior knowledge and regardless of the stability of the solutions.

While much progress in fluid dynamics has been made on the study of hydrodynamic instabilities, instabilities that occur when a control parameter is varied within the turbulent regime remain poorly understood \cite{fauveetal17}. The difficulty in studying such transitions arises from the underlying turbulent fluctuations which make analytical approaches cumbersome. These types of transitions resemble more closely phase transitions in statistical mechanics because the instability occurs on a fluctuating background \cite{kadanoff00}. In Rayleigh--B\'enard convection, such transitions have been observed in the form of reversals of the large scale circulation in an enclosed rectangular geometry, both in experiments and numerical simulations \cite{sugiyamaetal10,chandraverma13,podvinsergent17}. Although we do not study the turbulent case, we investigate the bifurcations of Rayleigh--B\'enard convection in an enclosed rectangular geometry to attempt to understand the genesis of the flow patterns that play the central role in flow reversals in the turbulent regime.

The paper is organised as follows. We first discuss the problem set-up and the choice of parameters in \cref{sec_problem}. Then, in \cref{sec_numerics}, we briefly review deflation and present the methods used to improve its initialisation procedure and compute the linear stability of the solutions. Next, we analyse the different branches discovered by deflation and their stability in \cref{sec_results}. Finally, in \cref{sec_conclusions}, we recapitulate the main conclusions of this study and propose potential extensions for future work.

\section{Problem set-up} \label{sec_problem}

We consider Rayleigh--B\'enard convection \cite{benard1900,rayleigh1916,benard1927} in a confined fluid heated from below and maintained at a constant temperature difference $\Delta T = T - T_0$ across a two dimensional square cell of height $d$. For simplicity, we employ the Oberbeck--Boussinesq approximation~\cite{oberbeck1879,boussinesq1903,tritton12} in which the kinematic viscosity $\nu$ and the thermal diffusivity $\kappa$ do not depend on temperature while the fluid density $\rho$ is assumed to be constant except in the buoyancy term of the momentum equation. In this case, we assume the density to depend linearly on the temperature, $\rho(T) \simeq \rho(T_0)[1 - \alpha\Delta T]$, where $\alpha$ is the thermal expansion coefficient. Then, the governing equations of the problem are
\begin{subequations} \label{eq_RB}
\begin{align}
\nabla \cdot \vf{u} &= 0,\\
\pd_t \vf{u} + \vf{u} \cdot \nabla \vf{u} &= - \nabla p + Pr \nabla^2 \vf{u} + Pr Ra T \hat{\vf{z}},\\
\pd_t T + \vf{u} \cdot \nabla T &= \nabla^2 T,
\end{align}
\end{subequations}
where $\vf{u}=(u,w)$ is the velocity field, $T$ is the temperature field, $p$ is the pressure, and $\hat{\vf{z}}$ is the buoyancy direction. The above equations have been nondimensionalized using $d$, $d^2/\kappa$ and $\Delta T$ as the relevant scales for length, time and temperature, respectively. The two dimensionless parameters of the problem are the Rayleigh and Prandtl numbers 
\begin{equation}
 Ra = g \alpha \Delta T d^3/\nu\kappa, \quad Pr = \nu / \kappa,  
\end{equation}
where $g$ is the gravitational acceleration. The Rayleigh number represents the ratio of the acceleration $g \alpha \Delta T$ related to buoyancy to the stabilizing effects of $\nu$ and $\kappa$. The bifurcation parameter in the problem we study is $Ra$ by fixing $Pr = 1$. The cell is assumed to have rigid walls, with thermally conducting horizontal walls and thermally insulating side walls, i.e.
\begin{subequations}
\begin{align}
 u = w &= 0, \quad x =0,1 \text{ or } z = 0, 1, \\
 T &= 1, \quad z = 0, \\
 T &= 0, \quad z = 1, \\
 \pd_xT &= 0, \quad x = 0, 1.
\end{align}
\end{subequations}

The trivial steady state is motionless with a negative thermal gradient through the layer,
\begin{equation} 
\label{eq:ststate}
 \vf{u} = 0, \qquad T = 1 - z,
\end{equation}
and is called the conducting state. This is because the fluid acts as a conducting material. The cooler fluid near the top of the layer is denser than the warmer fluid underneath it. 
The symmetries of the problem are: a) the mirror symmetry with respect to the axis $x = 1/2$
\begin{equation}
 [u,w,T](x,z) \to [-u,w,T](1-x,z),
 \label{eq:s1}
\end{equation}
which leaves the velocity and temperature field invariant, and b) the Boussinesq symmetry
\begin{equation}
 [u,w,T](x,z) \to [u,-w,1-T](x,1-z),
 \label{eq:s2}
\end{equation}
which leaves the velocity field invariant but transforms the temperature field into its opposite. These symmetries (and the combination of both) dictate the bifurcations that the system can undergo.

\section{Numerical methods} \label{sec_numerics}

\subsection{Computation of multiple solutions with deflation}
\label{sec_DCM}

Steady states of Rayleigh--B\'enard convection are computed using a recent numerical technique called deflation~\cite{farrell2015deflation}. This algorithm is based on Newton's method and can compute multiple solutions to a nonlinear system of equations $F(\phi,\lambda)=0$, where $\phi$ is the solution and $\lambda\in\mathbb{R}$ is a bifurcation parameter. First, fix $\lambda\in\mathbb{R}$ and assume a solution $\phi_1$ to $F(\cdot,\lambda)$ has been previously discovered by Newton's method. One can then construct a deflated problem
\begin{equation}
G(\phi,\lambda):=\left(\frac{1}{\|\phi-\phi_1\|^2}+1\right)F(\phi,\lambda),
\end{equation}
such that $G(\phi,\lambda)$ does not converge to zero as $\phi\to\phi_1$. The term $\mathcal{M}(\phi,\phi_1):= \|\phi-\phi_1\|^{-2}+1$ is called a deflation operator and ensures that Newton's method applied again to $G$ will not converge to a previously computed solution. By adding one to the expression of the deflation operator, we impose that the new problem $G$ behaves similarly to $F$ away from the root $\phi_1$. This process can be repeated to obtain a set of solutions $S(\lambda)$ to a nonlinear problem $F(\phi,\lambda)=0$ at the bifurcation parameter $\lambda$. Deflated continuation is the combination of this idea with continuation. The previously discovered solutions are continued to a new bifurcation parameter $\lambda-\Delta \lambda$: each $\phi\in S(\lambda)$ is used as an initial guess for the deflation procedure in order to construct a new set of solutions $S(\lambda-\Delta\lambda)$ at $\lambda-\Delta \lambda$.

We apply this algorithm repeatedly to discover multiple solutions to the steady state Rayleigh--B\'enard problem:
\begin{equation} \label{eq_steady_RB}
F(\phi, Ra):=
\begin{cases}
- \vf{u} \cdot \nabla \vf{u} - \nabla p + Pr \nabla^2 \vf{u} + Pr Ra T \hat{\vf{z}} &=0,\\
\nabla \cdot \vf{u} &= 0,\\
- \vf{u} \cdot \nabla T + \nabla^2 T &=0,
\end{cases}
\end{equation}
where $\phi=(\vf{u},p,T)$ and $Ra$ is the bifurcation parameter. This equation is discretised with Taylor--Hood finite elements for the velocity and pressure (piecewise biquadratic and piecewise bilinear respectively), together with piecewise bilinear polynomials for the temperature, using the Firedrake finite element library~\cite{rathgeber2016}.
The unit square domain is represented by a mesh with $50 \times 50$ square cells to preserve the symmetries of the problem and avoid spurious symmetry breaking solutions. A coarse discretization is needed due to the complexity of the problem and number of distinct solutions discovered. However, we did not observe significant differences in selected experiments where we refined the mesh.

Deflated continuation requires two different computational tasks. The first is the continuation of a previously discovered branch by the continuation step, which usually results in the convergence of the Newton solver in an average of three iterations. The convergence criterion is defined to be when the Euclidean $l^2$-norm of the discretised PDE residual is below $10^{-8}$. The second task is the deflation step to discover new branches. Most of the time, the solver does not converge, which means that it does not discover any new branch within the maximum number of Newton iterations allowed, which is set to one hundred. Typically, when the deflation step is successful, it requires around $80$ iterations to find a new solution. Since the deflation step is employed for each of the steady states in the different branches, this yields at least hundreds of thousands of Newton iterations in total, and a few weeks of calculation on a desktop with 32 cores. The linear systems resulting from Newton's method are solved using the sparse direct solver MUMPS~\cite{amestoy2000mumps}.

A crucial question in the deflation technique is the choice of the deflation operator $\mathcal{M}$ because it can affect the convergence of Newton's method. We use the following operator,
\begin{equation}
\mathcal{M}((\vf{u},p,T),(\vf{u}_1,p_1,T_1))=\left(\frac{1}{\|\vf{u}-\vf{u}_1\|_2^2+\|\nabla(\vf{u}-\vf{u}_1)\|_2^2+\|T-T_1\|_2^2}+1\right),
\end{equation}
where $\|\cdot\|_2$ denotes the $L^2$-norm. Note that the deflation operator $\mathcal{M}$ does not depend on the pressure $p$ because we wish to deflate all the solutions of the form $(u_1,p_1+c,T_1)$, where $c\in\mathbb{R}$, to avoid discovering solutions trivially related to known ones.

We perform bifurcation analysis of Rayleigh--B\'enard convection and compute solutions to the nonlinear system of \cref{eq_steady_RB} for $0 \leq Ra \leq 10^5$ using deflated continuation for $Ra=10^5,10^5-\Delta R_a,\ldots,0$. Here, $\Delta Ra$ denotes the continuation step size in the bifurcation parameter, which is chosen as $\Delta Ra = 100$. In a previous work on the computation of solutions to the 3D nonlinear Schr\"odinger equation~\cite{boulle2020deflation}, we found that having good initial guesses (e.g.~solutions of the linearised problem) for the initial deflation steps facilitates the convergence of Newton's method and leads to more complex and interesting states. However, contrary to~\cite{boulle2020deflation} we do not have the analytic expressions of the linear solutions to the Rayleigh--B\'enard convection due to the no-slip boundary conditions. Therefore, we numerically solve an eigenvalue problem (see \cref{sec:stability}) to obtain the first ten unstable eigenmodes $(\vf{u}_1,T_1),\ldots,(\vf{u}_{10},T_{10})$ linearised around the conducting steady state \cref{eq:ststate} (see \cref{fig:eigenmodes}). Then, the sums of the conducting state with the respective normalized perturbations: $(\vf{u}_1,1-z+T_1),\ldots,(\vf{u}_{10},1-z+T_{10})$ are used as initial guesses for deflated continuation.

\subsection{Linear stability analysis} 
\label{sec:stability}

The stability analysis of a given steady state $(\vf{u}_0,T_0)$ is performed by considering the perturbation ansatz
\begin{subequations} \label{eq_linear_ansatz}
\begin{align}
\vf{u}(x,z,t) &= \vf{u}_0 + \vf{v} e^{\lambda t}, \\ %\epsilon \vf{u}e^{\lambda t},\\
T(x,z,t) &= T_0 + \theta e^{\lambda t}, %\epsilon Te^{\lambda t},
\end{align}
\end{subequations}
where the velocity perturbation $\vf{v} \ll 1$ and the temperature perturbation $\theta \ll 1$ have an eigenvalue $\lambda$, whose real part is the growth rate and imaginary part is the frequency of the corresponding eigenvector. When at least one of the eigenvalues has a positive real part $\mathcal R(\lambda) > 0$, we can have two types of instabilities: to a real eigenmode, occurring when the imaginary part of the eigenvalue $\mathcal I(\lambda) = 0$, and to a complex eigenmode, occurring when $\mathcal I(\lambda) \neq 0$. If all the eigenvalues have negative real parts, then the steady state is stable. Inserting \cref{eq_linear_ansatz} into \cref{eq_RB} and considering the linearised system of equations, we obtain the following generalised eigenvalue problem at leading order
\begin{equation}
\begin{pmatrix} \label{eq_eigenvalue_pb}
A & -\nabla & RaP_r\hat{\vf{z}}\\
\nabla\cdot & 0 & 0\\
-\nabla T_0\cdot & 0 & \nabla^2-\vf{u}_0\cdot\nabla
\end{pmatrix}
\begin{pmatrix}
\vf{v}\\ p\\ \theta
\end{pmatrix}=
\lambda
\begin{pmatrix}
I & 0 & 0\\
0 & 0 & 0\\
0 & 0 & I
\end{pmatrix}
\begin{pmatrix}
\vf{v}\\ p\\ \theta
\end{pmatrix},
\end{equation}
where $I$ is the relevant identity operator and $A$ is the linear operator defined by
\[A\vf{v}=\nabla^2\vf{v}-\vf{u}_0\cdot\nabla\vf{v}-\vf{v}\cdot\nabla\vf{u}_0.\]
The eigenvalue problem described in \cref{eq_eigenvalue_pb} is solved with a Krylov--Schur method~\cite{stewart2002} using the SLEPc library~\cite{hernandez2005slepc}. 

To obtain the critical value of the Rayleigh number $Ra_c$ at which the conducting state \cref{eq:ststate} becomes unstable, we modify the eigenvalue problem in \cref{eq_eigenvalue_pb}. Inserting \cref{eq:ststate} into \cref{eq_eigenvalue_pb} we obtain the following generalized eigenvalue problem
\begin{equation}
\begin{pmatrix} \label{eq_eigenvalue_pb_motionless}
\nabla^2 & -\nabla & RaP_r\hat{\vf{z}}\\
\nabla\cdot & 0 & 0\\
\hat{\vf{z}}\cdot & 0 & \nabla^2
\end{pmatrix}
\begin{pmatrix}
\vf{v}\\ p\\ \theta
\end{pmatrix}=
\lambda
\begin{pmatrix}
I & 0 & 0\\
0 & 0 & 0\\
0 & 0 & I
\end{pmatrix}
\begin{pmatrix}
\vf{v}\\ p\\ \theta
\end{pmatrix}.
\end{equation}
A steady state becomes unstable when $\lambda=0$ and this corresponds to the critical Rayleigh number $Ra_c$ which is the smallest $Ra$ satisfying \cref{eq_eigenvalue_pb_motionless} for $\lambda=0$. So, to obtain the critical value of the Rayleigh number for different base states we have reformulated the problem to a generalised eigenvalue problem for $Ra_c$ as follows
\begin{equation}
\begin{pmatrix} \label{eq_eigenvalue_pb_rayleigh}
\nabla^2 & -\nabla & 0\\
\nabla\cdot & 0 & 0\\
\hat{\vf{z}}\cdot & 0 & \nabla^2
\end{pmatrix}
\begin{pmatrix}
\vf{v}\\ p\\ \theta
\end{pmatrix}=
Ra_c
\begin{pmatrix}
0 & 0 & -P_r\hat{\vf{z}}\\
0 & 0 & 0\\
0 & 0 & 0
\end{pmatrix}
\begin{pmatrix}
\vf{v}\\ p\\ \theta
\end{pmatrix}.
\end{equation}
In the following sections this linear analysis will be used to study the stability of the conducting state \cref{eq:ststate} and of the nonlinear steady states that we obtain from deflated continuation.

\section{Results} \label{sec_results}

\subsection{Primary instabilities} \label{sec_primary_instabilities}

\begin{figure}[!ht]
\centering
\includegraphics[width=\figwidth]{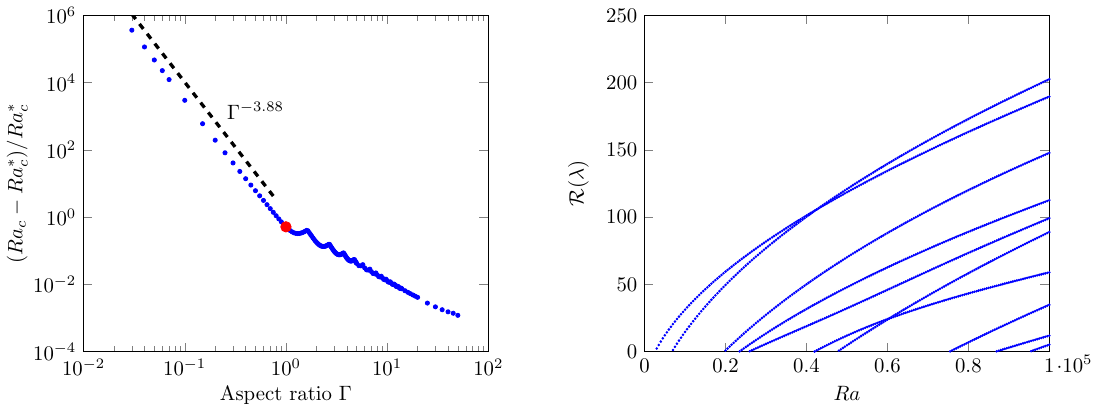}
\caption{(Color online) Left: Critical Rayleigh number with respect to the aspect ratio $\Gamma$ (blue dots). The red dot indicates the aspect ratio $\Gamma=1$, which is the focus of our study. Right: Growth rates of the eigenmodes that emanate from the conducting steady state.}
\label{fig:critical_ratio}
\end{figure}

In this section, we analyse the stability of the conducting steady state defined by \cref{eq:ststate}. We found that the first instability of this state arises at $Ra_c^{(1)} := Ra_c\approx 2586$. It is well known~\cite{chandrasekhar1961hydrodynamic} that the critical Rayleigh number is approximately equal to $Ra_c^*\approx 1707.762$ for a domain with periodic side boundaries and no-slip top and bottom boundaries. This difference in the value of $Ra_c$ is due to the effect of the side walls on the instability. This effect should decrease when the length of the domain, $L$, becomes large such that the behaviour predicted for an unbounded domain is recovered. To demonstrate this, we systematically increase the aspect ratio
\begin{equation}
 \Gamma = \frac{L}{d}.
 \label{eq:gamma}
\end{equation}
Figure \ref{fig:critical_ratio} (left) clearly shows that for $\Gamma \gg 1$ the $Ra_c^{(1)}$ converges to $Ra_c^*$. 
In addition, for aspect ratios $\Gamma < 1$ we observe $Ra_c \propto \Gamma^{-3.88}$. The critical Rayleigh number for Rayleigh--B\'enard convection with stress-free boundary conditions is $Ra_c^*(k) = (\pi^2 + k^2)^3/k^2$ \cite{fauve17}, where $k$ is the horizontal wavenumber. For $k \gg 1$ the critical Rayleigh number scales like $Ra_c^*(k) \propto k^4  \propto \Gamma^{-4}$ using \cref{eq:gamma}. This scaling relation is very close to the observed one, which implies that the exponent is not affected much by the no-slip boundary conditions. The exponent we find is in agreement with \cite{mizushima95}.

\begin{table}[!ht]
\caption{The ten primary bifurcations from the conductive state within the range $0 \leq Ra \leq 10^5$. Here $n$ enumerates the eigenmodes by order of appearance as $Ra$ increases. Here $(m_x,m_z)$ are the number of rolls in the horizontal and vertical directions.}
\label{tbl:modes}
\begin{center}
\begin{tabular}{c|*{10}{c}}
\hline
\hline
$n$ & 1 & 2 & 3 & 4 & 5 & 6 & 7 & 8 & 9 & 10 \\ 
$Ra_c^{(n)}$ & 2586 & 6746 & 19655 & 23346 & 25780 & 41772 & 47431 & 74878 & 86313 & 94543 \\ 
$(m_x,m_z)$ & (1,1) & (2,1) & (3,1) & (2,2) & (1,2) & (3,2) & (4,1) & (4,2) & (2,3) & (3,3) \\ 
\hline
\hline
\end{tabular}
\end{center}
\end{table}

The case on which we focus for the rest of the paper is the square cell ($\Gamma = 1$), which is denoted with a red dot in \cref{fig:critical_ratio} (left). We first study the linear stability of the conducting state \cref{eq:ststate}. In the range $0 \leq Ra \leq 10^5$ we observe 10 supercritical stationary bifurcations that arise from the conducting state. Figure \ref{fig:critical_ratio} (right) shows the growth rates of the unstable eigenmodes as a function of $Ra$. These bifurcations occur when $\mathcal R(\lambda) = 0$. The critical values of the Rayleigh number at the onsets are listed in \cref{tbl:modes}. \cref{fig:critical_ratio} (right) demonstrates that the growth rates of the eigenmodes vary with Rayleigh number such that the curves can cross. The eigenmode which bifurcated from the conducting state at $Ra_c^{(2)} = 6746$ ends up being the most unstable as $Ra \to 10^5$. 
\begin{figure}[!ht]
\centering
$Ra_c^{(1)} = 2586$ \hspace{0.5cm}
$Ra_c^{(2)} = 6746$ \hspace{0.55cm}
$Ra_c^{(3)} = 19655$ \hspace{0.5cm}
$Ra_c^{(4)} = 23346$ \hspace{0.5cm}
$Ra_c^{(5)} = 25780$ \\
\vspace{0.2cm}
\includegraphics[width=0.11\textwidth]{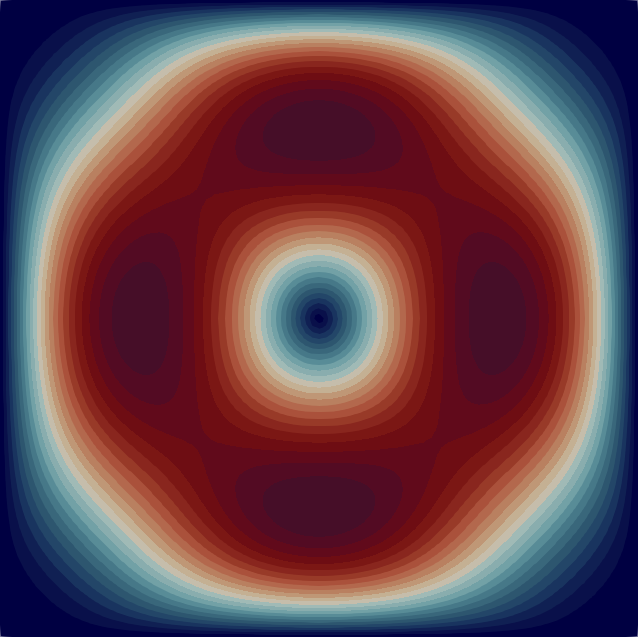}\hspace{0.6cm}
\includegraphics[width=0.11\textwidth]{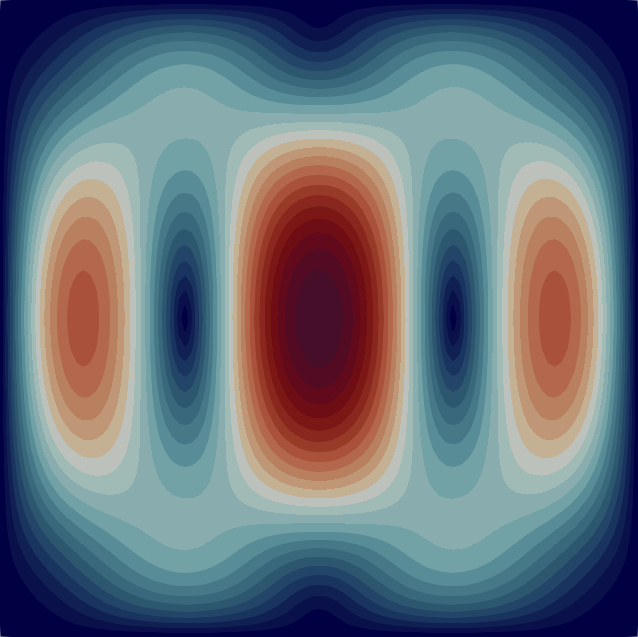}\hspace{0.6cm}
\includegraphics[width=0.11\textwidth]{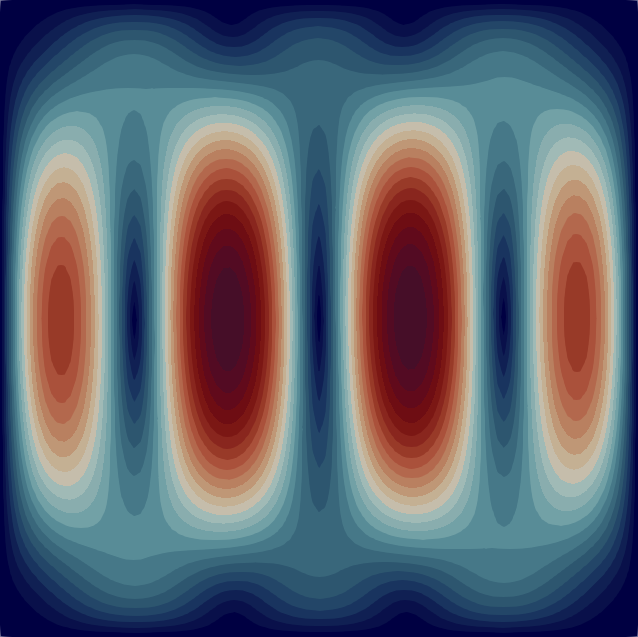}\hspace{0.6cm}
\includegraphics[width=0.11\textwidth]{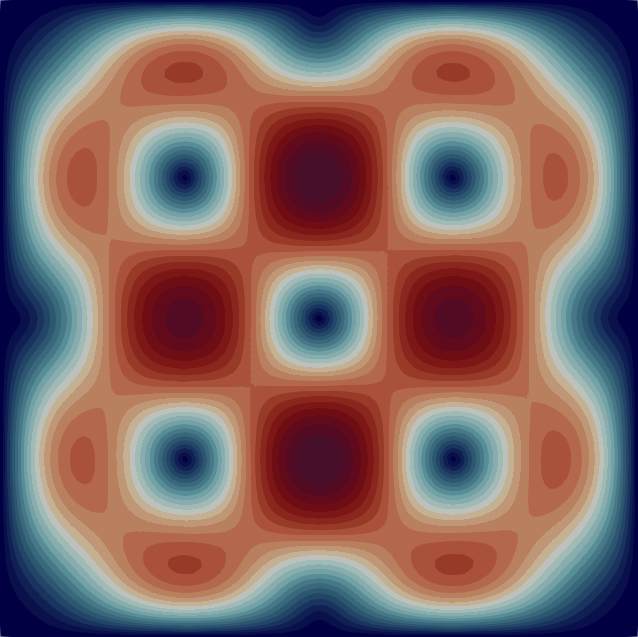}\hspace{0.6cm}
\includegraphics[width=0.11\textwidth]{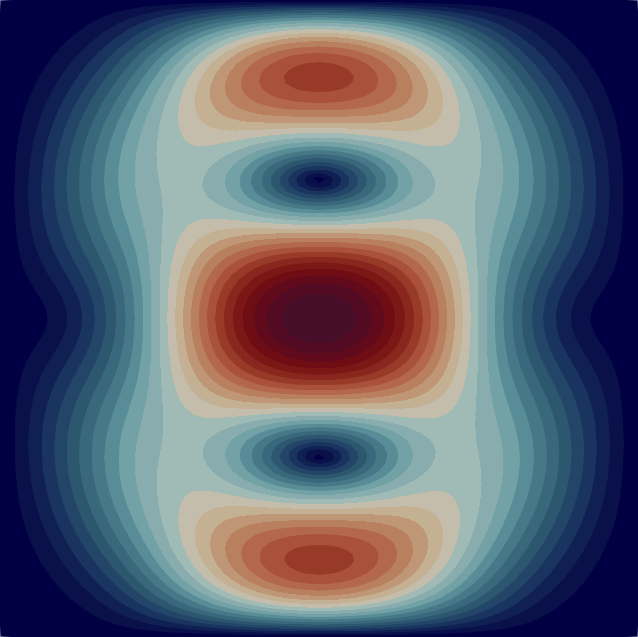}
\\
\vspace{0.3cm}
\includegraphics[width=0.11\textwidth]{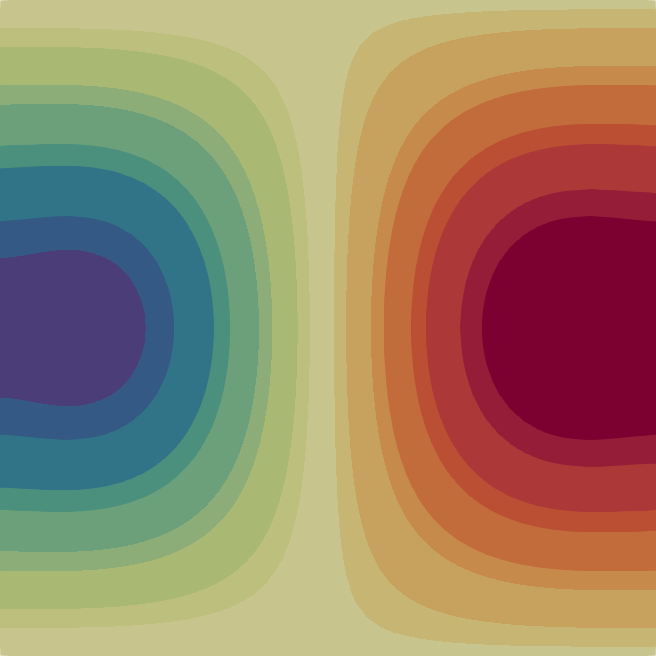}\hspace{0.6cm}
\includegraphics[width=0.11\textwidth]{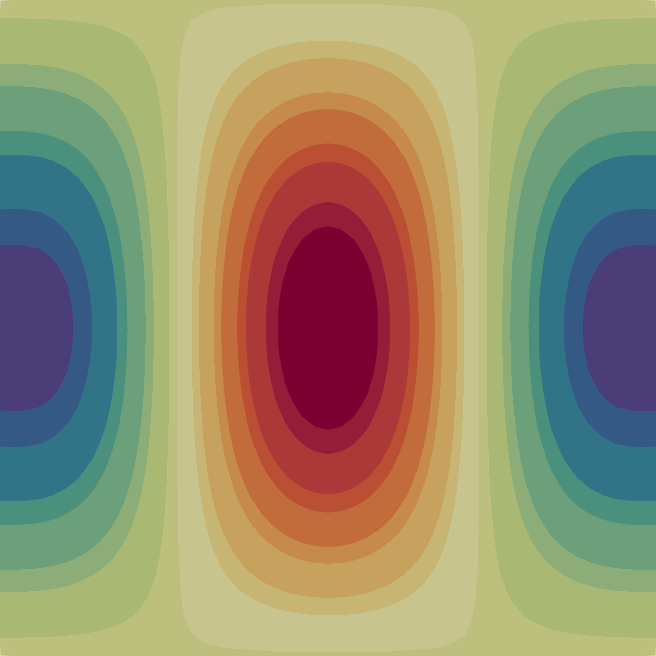}\hspace{0.6cm}
\includegraphics[width=0.11\textwidth]{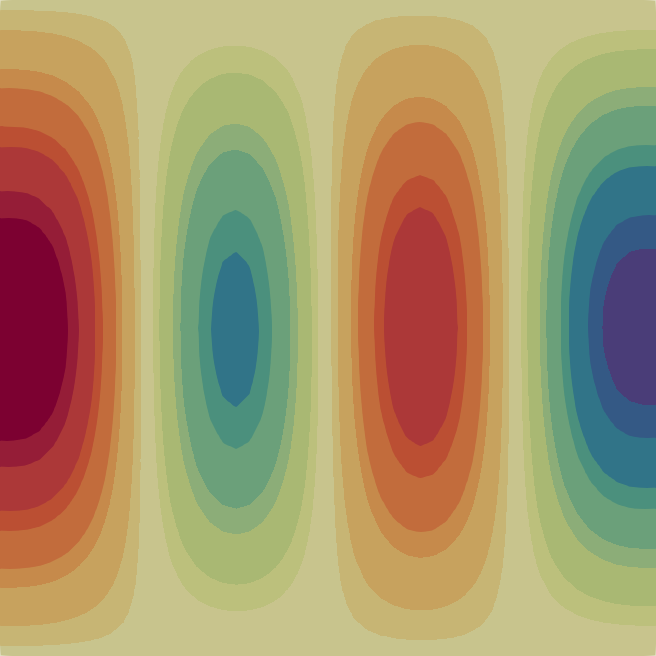}\hspace{0.6cm}
\includegraphics[width=0.11\textwidth]{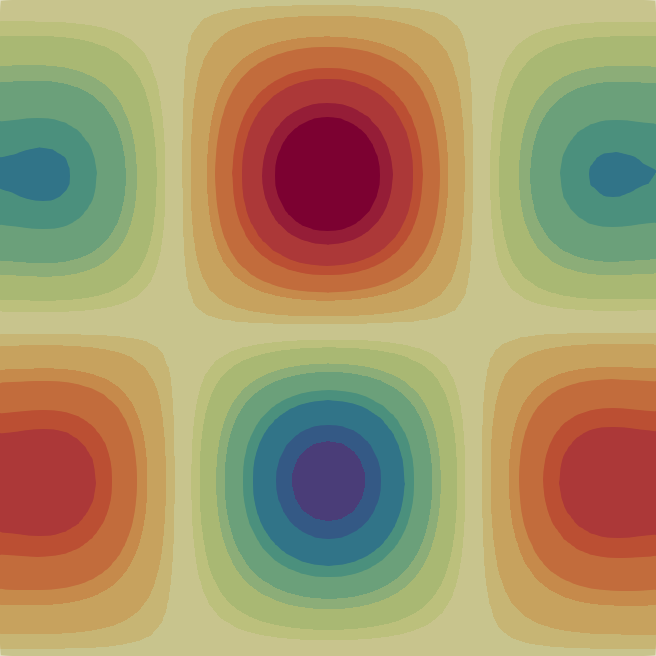}\hspace{0.6cm}
\includegraphics[width=0.11\textwidth]{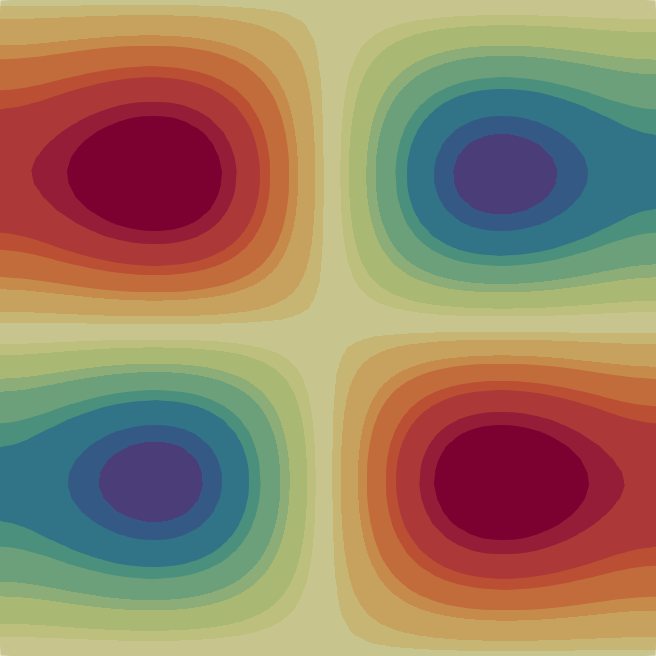}
\\
\vspace{0.6cm}
$Ra_c^{(6)} = 41772$ \hspace{0.4cm}
$Ra_c^{(7)} = 47431$ \hspace{0.4cm}
$Ra_c^{(8)} = 74878$ \hspace{0.4cm}
$Ra_c^{(9)} = 86313$ \hspace{0.4cm}
$Ra_c^{(10)} = 94543$ \\
\vspace{0.2cm}
\includegraphics[width=0.11\textwidth]{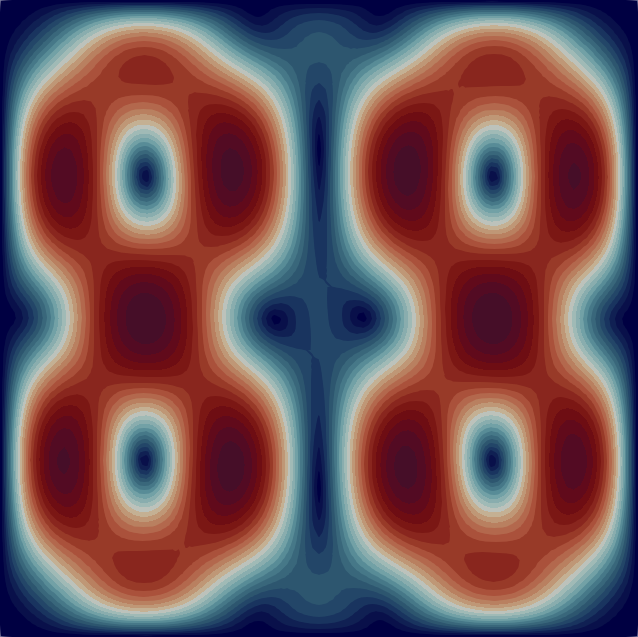}\hspace{0.6cm}
\includegraphics[width=0.11\textwidth]{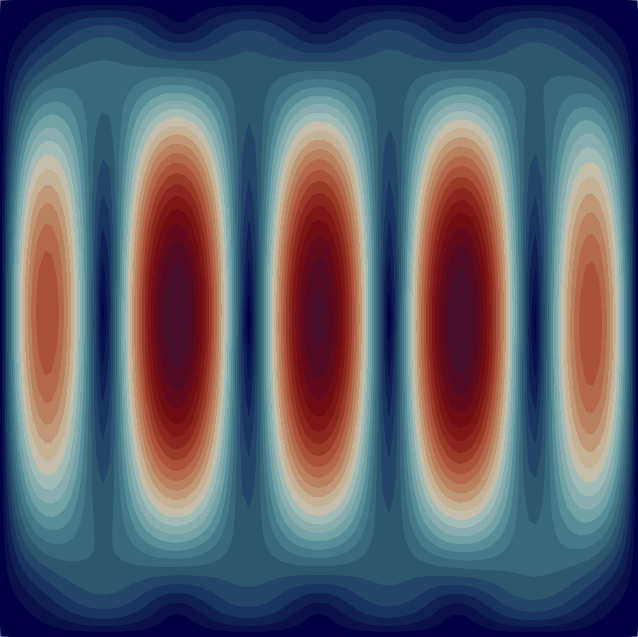}\hspace{0.6cm}
\includegraphics[width=0.11\textwidth]{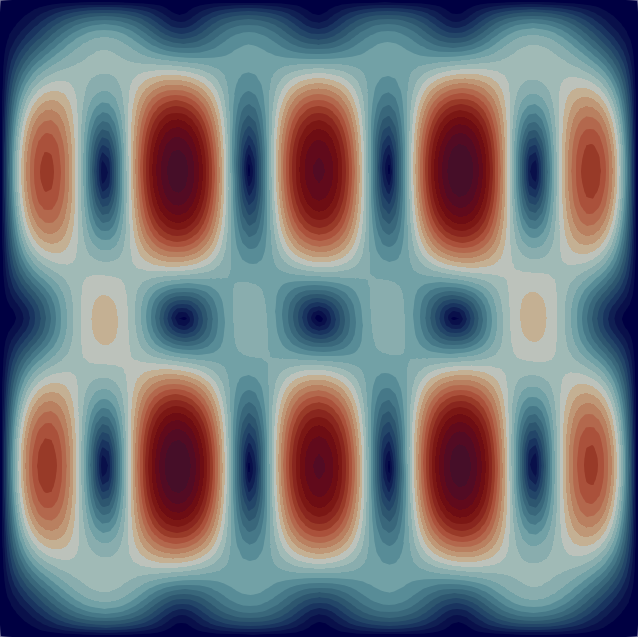}\hspace{0.6cm}
\includegraphics[width=0.11\textwidth]{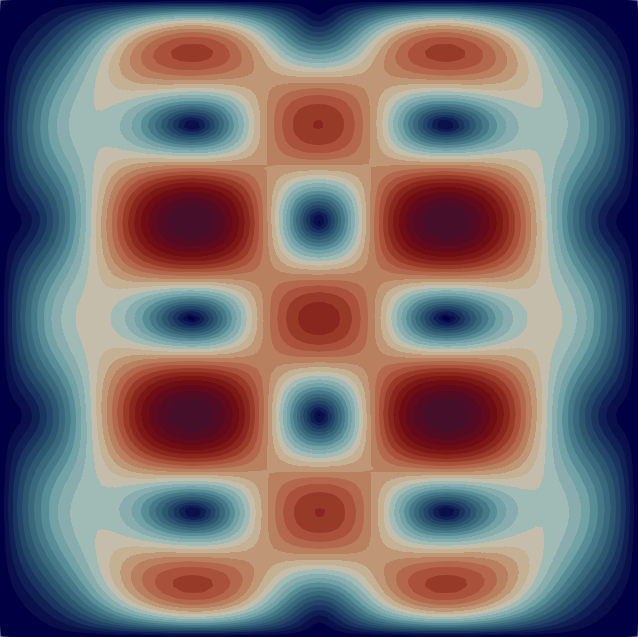}\hspace{0.6cm}
\includegraphics[width=0.11\textwidth]{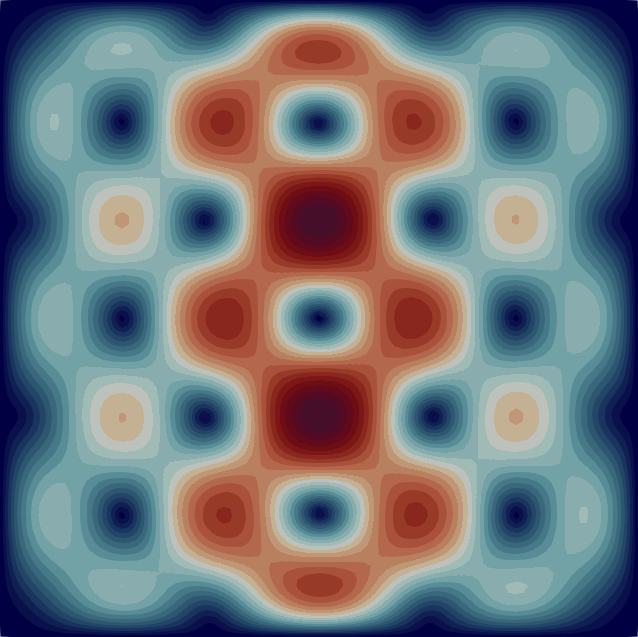}
\\
\vspace{0.3cm}
\includegraphics[width=0.11\textwidth]{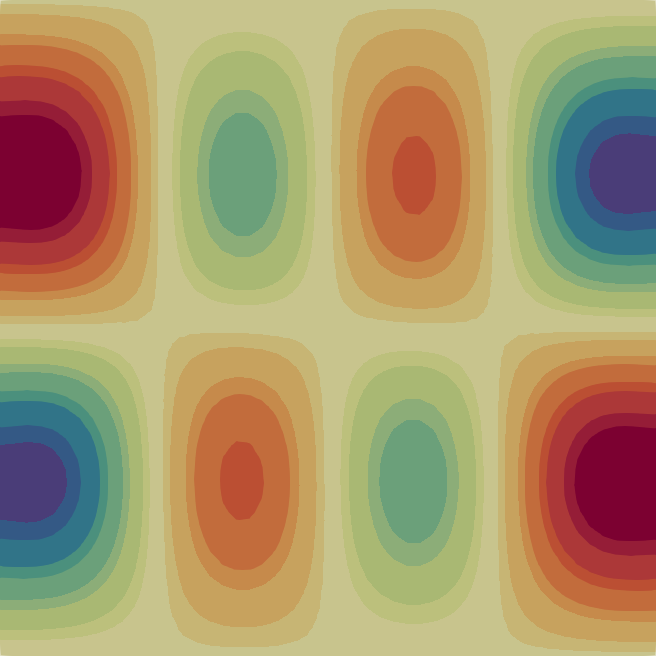}\hspace{0.6cm}
\includegraphics[width=0.11\textwidth]{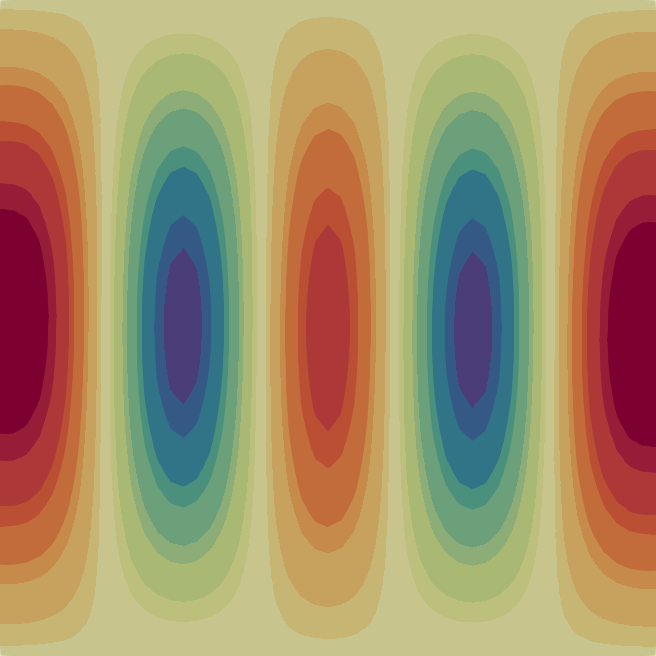}\hspace{0.6cm}
\includegraphics[width=0.11\textwidth]{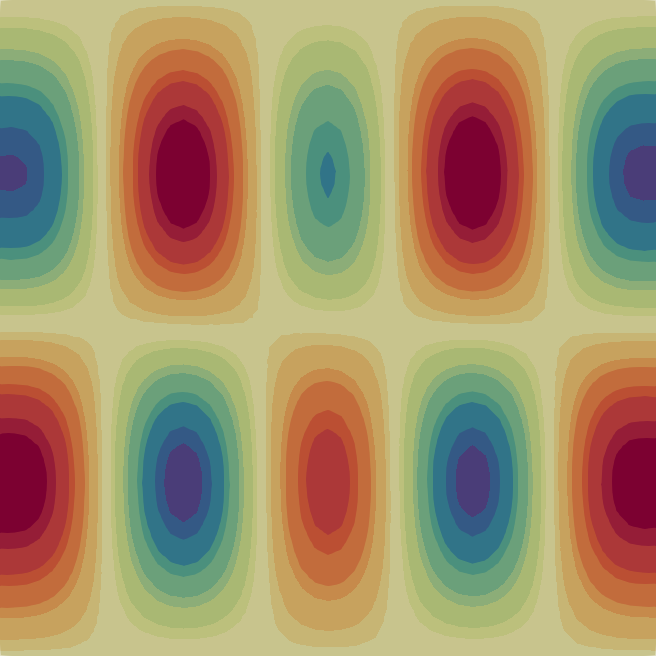}\hspace{0.6cm}
\includegraphics[width=0.11\textwidth]{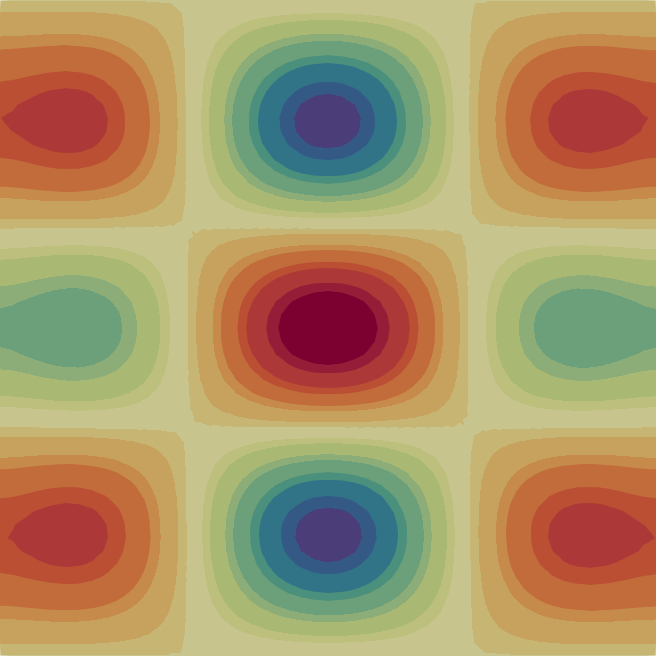}\hspace{0.6cm}
\includegraphics[width=0.11\textwidth]{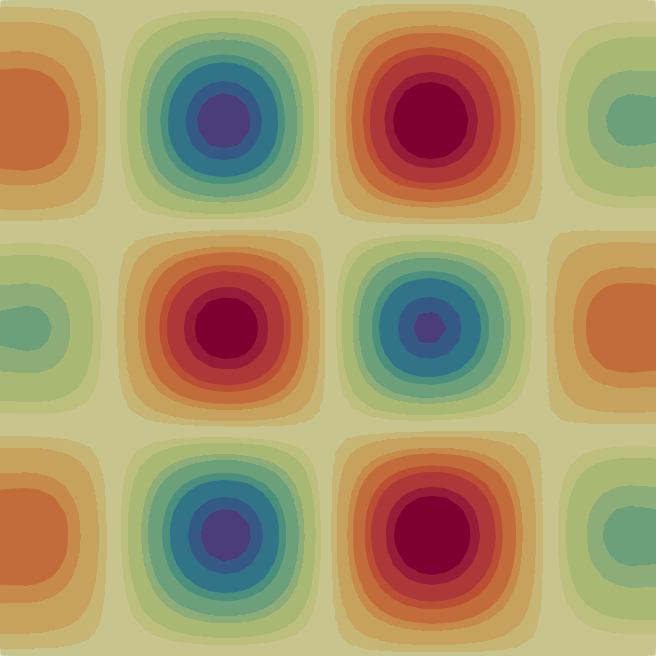}
\caption{(Color online) The eigenmodes of the primary bifurcations emanating from the conducting steady state Eq. \eqref{eq:ststate} in the Rayleigh number range $0 \leq Ra \leq 10^5$. The top rows show the magnitude of the velocity $\vf{v}$ (red color indicates a high magnitude and blue a zero velocity magnitude), while the bottom rows show the temperature $\theta$ (red and blue respectively display positive and negative temperature values).}
\label{fig:eigenmodes}
\end{figure}
The velocity magnitude and temperature fields of the eigenmodes from these primary bifurcations of the flow are shown in \cref{fig:eigenmodes} with the corresponding critical Rayleigh numbers indicated on the top of each flow pattern. In \cref{tbl:modes} we characterise the flow patterns by the number of rolls $(m_x,m_z)$, where $m_x$ and $m_z$ are the number of rolls in the temperature field in the horizontal and vertical direction, respectively. Note that due to the aspect ratio $\Gamma =1$ of the domain the flow pattern of the first bifurcation at $Ra_c^{(1)} = 2586$ is just a single convection roll (see \cref{fig:eigenmodes}) according to the classical result \cite{rayleigh1916}. In the following sections, we first analyse the branches of solutions that are created from some of these primary bifurcations before moving to disconnected branches.

\subsection{Bifurcation diagrams}

The combination of deflation and the initialisation strategy presented in \cref{sec_DCM} identified $43$ branches for the two-dimensional steady-state Rayleigh--B\'enard convection at $Ra=10^5$, leading to $129$ branches when reflections are included. Here, we analyse the evolution and the linear stability of some of the branches arising from the first four excited states (see~\cref{fig:eigenmodes}) as well as two disconnected branches. Moreover, we discard the mirror symmetric solutions with respect to the $x=1/2$ and $z=1/2$ axes (see symmetries in \cref{sec_problem}). 
Our diagnostics for the bifurcation diagrams, representing the evolution of the steady states as a function of $Ra$, are the kinetic energy $\|\vf{u}\|_2^2$, the potential energy $\|T\|_2^2$, and the Nusselt number $Nu$, which are defined as
\begin{equation}
\|\vf{u}\|_2^2=\int \vf{u}^2 d^2x, \qquad \|T\|_2^2=\int T^2 d^2x, \qquad Nu = \int |\nabla T|^2 d^2x.
\end{equation}
The Nusselt number characterises the efficiency of convective heat transfer and is given by the ratio of the total heat transfer (i.e.~both advective and diffusive) to the conductive heat transfer $Nu \equiv H L / \kappa d = \avg{w T} - \kappa \pd_z\avg{T} = \avg{(\nabla T)^2}$, where $H$ is the heat flux and $\avg{\cdot}$ stands for the area average. A Nusselt number of one represents heat transfer by pure conduction. Here, the Nusselt number is essentially defined as the dissipation rate of the temperature variance (see second equality). This is obtained by integrating the equation for potential energy over the entire area \cite{siggia94}.

\begin{figure}[!ht]
\vspace{0.2cm}
\center
\begin{overpic}[width=\figwidth]{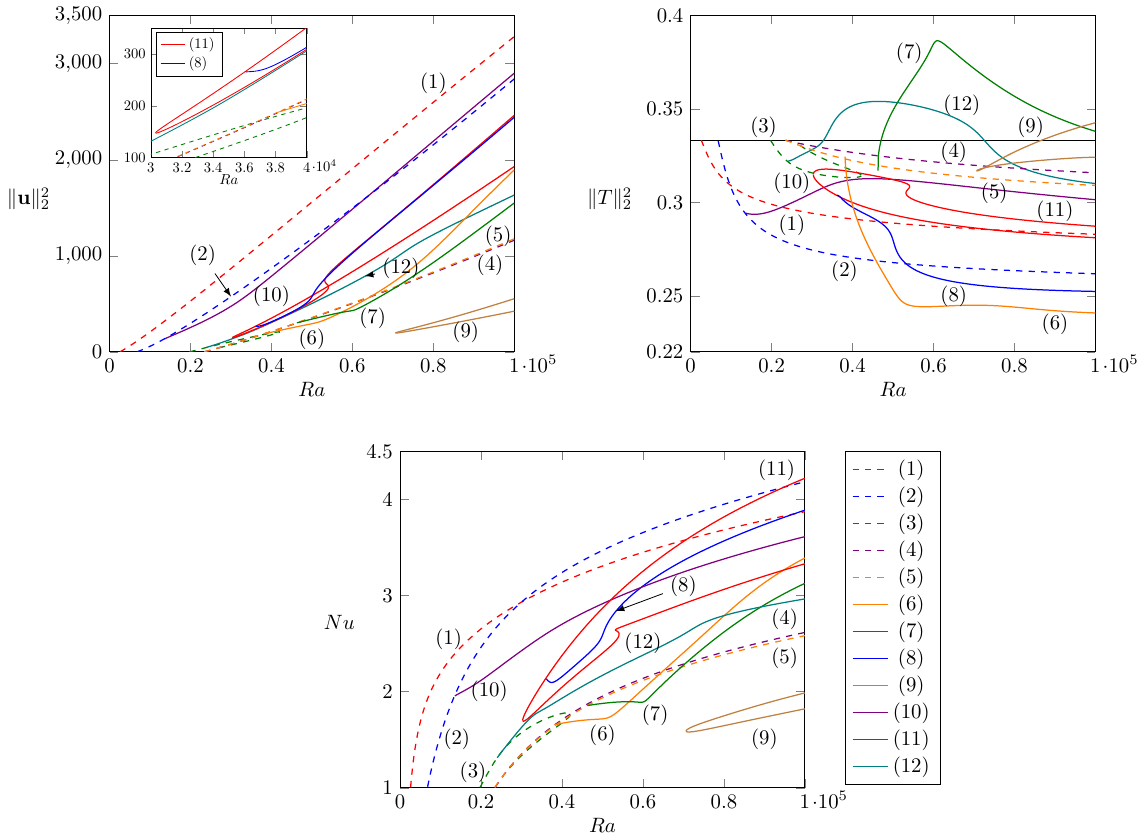}
\put(1,73.5){(a)}
\put(51.5,73.5){(b)}
\put(28,34){(c)}
\end{overpic}
\caption{(Color online) Bifurcation diagram using the kinetic energy $\|\vf{u}\|_2^2$ of the branches studied in the paper (a). The panels (b) and (c) represent the potential energy $\|T\|_2^2$ and the Nusselt number $Nu$ of the branches as a function of $Ra$, respectively. The branches that originate from the linear conducting state are plotted in dashed lines, while branches emanating from secondary instabilities and disconnected branches are depicted with solid lines. The horizontal black line in (b) indicates the potential energy of the conducting steady state.}
\label{fig:bif_diagrams}
\end{figure}

The bifurcation diagrams are presented in \cref{fig:bif_diagrams}, where the numbers denote the different branches of steady solutions, indicating the values of kinetic energy, potential energy and Nusselt number that correspond to each branch on each diagram. 
The first steady solutions that we obtain from deflated continuation as $Ra$ is varied are four branches that arise from the first four bifurcations of the conducting steady state: branch (1) at $Ra_c^{(1)}=2586$, branch (2) at $Ra_c^{(2)}=6746$, branch (3) at $Ra_c^{(3)}=19655$ and branch (4) at $Ra_c^{(4)}=23346$. Moreover a secondary branch (10) bifurcates at $Ra \approx 13550$ from branch (2), which originates from the second instability of the conducting state. A similar behaviour is observed for the fourth instability, where branch (6) bifurcates at $Ra \approx 27300$ from branch (4). We also focus on two disconnected branches (11) and (8) obtained directly by deflated continuation. In \cref{fig:bif_diagrams}(a), branch (11) is close to branch (12). However, note that the bifurcation diagram using the potential energy (see \cref{fig:bif_diagrams}(b)) shows that branch (11) does not bifurcate from branch (12). On the other hand, branch (8) bifurcates from branch (11) at $Ra \approx 37000$ (see \cref{fig:bif_diagrams}).
It is interesting to point out that in the range $0 \leq Ra \leq 10^5$ branch (1) has the highest kinetic energy while branch (6) has the highest potential energy (see \cref{fig:bif_diagrams}(a) and \ref{fig:bif_diagrams}(b)). The steady states that are most effective in convecting heat transfer for $0 \leq Ra \lesssim 3 \times 10^4$ are in branch (1) from the first instability of the conducting state while for $3 \times 10^4 \lesssim Ra \leq 9.5\times 10^4$ are in the branch (2), and in branch (11) for $Ra\gtrsim 9.5\times 10^4$ (see \cref{fig:bif_diagrams}(c)).

\subsection{Bifurcations from the conducting steady state \label{sec:eigenmodes}}
In this section we analyse the states that emanate from the conducting steady state and their evolution as a function of the bifurcation parameter $Ra$. Our analysis involves the detailed evolution of the velocity magnitude and the temperature on the branches of the bifurcation diagrams using the kinetic and potential energy, respectively. In addition, we present the largest growth rates and corresponding frequencies from the linear stability analysis we have performed on the steady states.

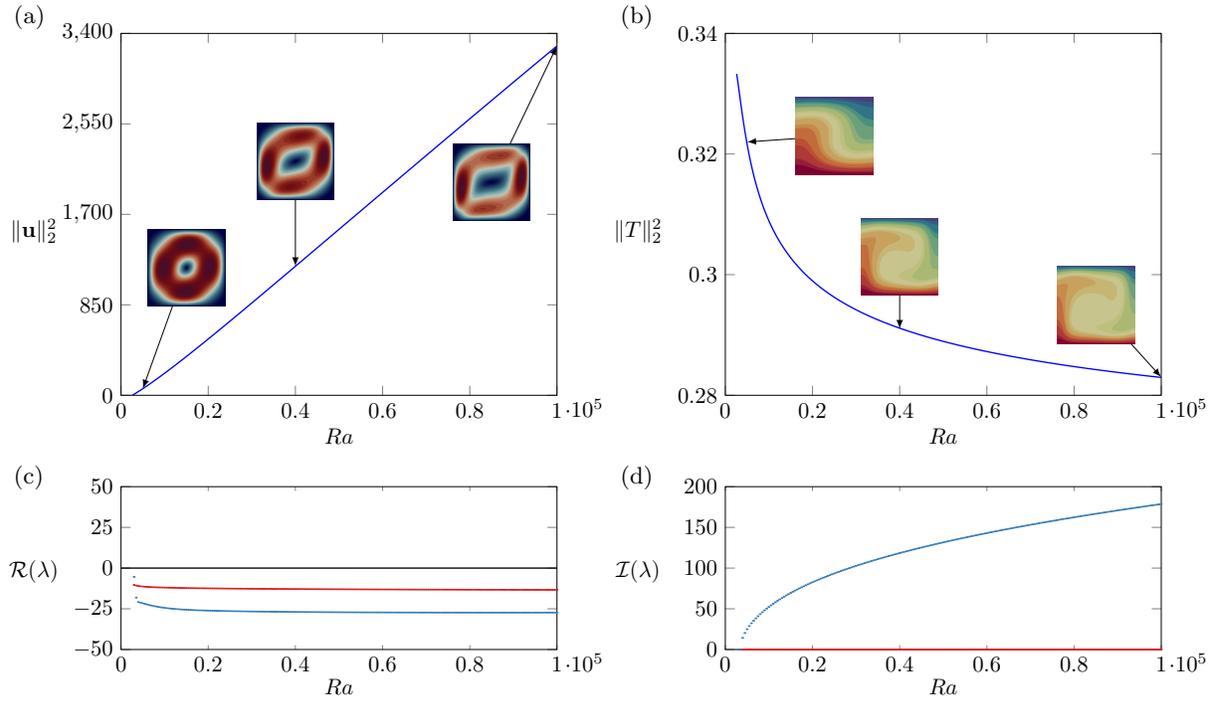
\begin{figure}[htbp]
\vspace{0.2cm}
\centering
\begin{overpic}[width=\figwidth]{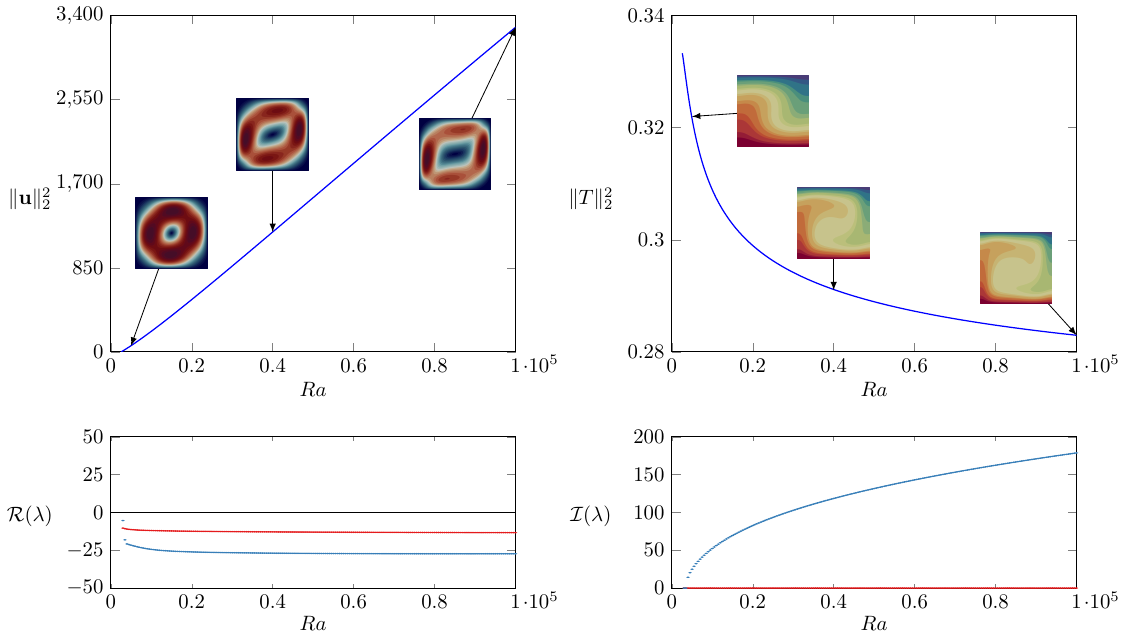}
\put(1,56){(a)}
\put(51,56){(b)}
\put(1,18){(c)}
\put(51,18){(d)}
\end{overpic}
\caption{(Color online) Evolution of the steady states in branch (1) arising from the first eigenmode, illustrated via (a) the kinetic energy and (b) the potential energy. The color scheme for the temperature ranges from 0 (blue) to 1 (red).  The largest growth rates and corresponding frequencies are presented in (c) and (d), respectively.}
\label{fig:branch_a}
\end{figure}

Figure \ref{fig:branch_a} shows results of the aforementioned analysis for branch (1), which arises at $Ra_c^{(1)}=2586$ from the first eigenmode (see \cref{fig:eigenmodes}). This bifurcation breaks the $x$-reflection symmetry but the branch has the Boussinesq symmetry. As $Ra$ increases, the magnitude of the velocity field evolves from a circular to a bent convection roll at $Ra \approx 4 \times 10^4$ (see \cref{fig:branch_a}(a)). On the other hand, the $z$-shaped interface of the temperature field is sheared (see \cref{fig:branch_a}(b)), enhancing the heat transfer in the convection cell (see also branch (1) in \cref{fig:bif_diagrams}(c)). \cref{fig:branch_a}(c) demonstrates that the solutions on this branch remain stable ($\mathcal R(\lambda) < 0$) with an almost constant growth rate over the whole range of Rayleigh numbers we consider. The frequency corresponding to the largest growth rate is zero from the stationary bifurcation at $Ra_c^{(1)}=2586$ to the $Ra = 10^5$ (see \cref{fig:branch_a}(d)), while the next eigenvalue has a negative real part and a large imaginary part.

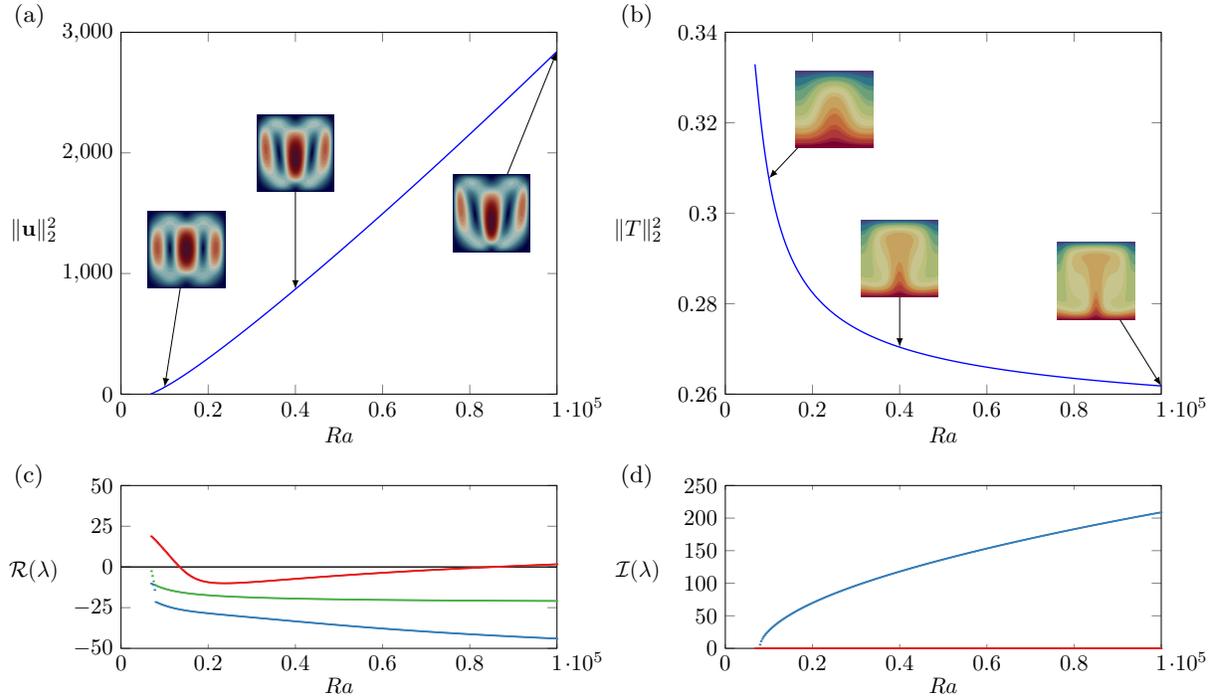
\begin{figure}[htbp]
\vspace{0.2cm}
\centering
\begin{overpic}[width=\figwidth]{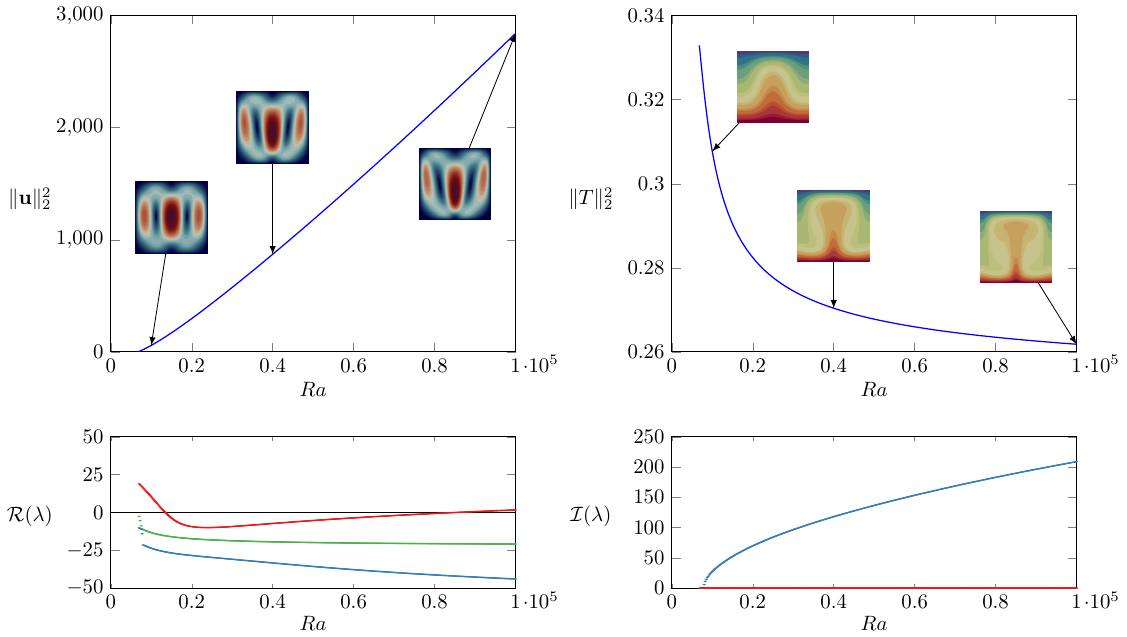}
\put(1,56){(a)}
\put(51,56){(b)}
\put(1,18){(c)}
\put(51,18){(d)}
\end{overpic}
\caption{(Color online) Evolution of the steady states in branch (2) arising from the second eigenmode, illustrated via (a) the kinetic energy and (b) the potential energy. The largest growth rates and corresponding frequencies are presented in (c) and (d), respectively. The two largest growth rates are shown in red and green while the subsequent ones are depicted in blue.}
\label{fig:branch_0}
\end{figure}

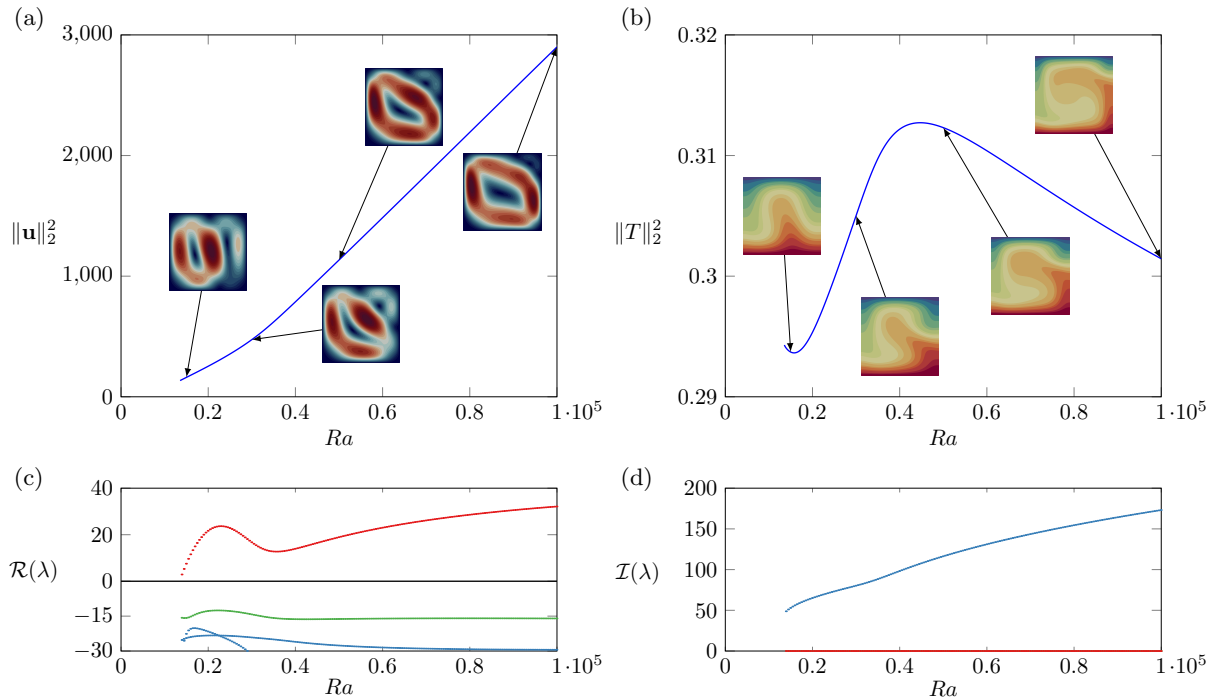
\begin{figure}[htbp]
\vspace{0.2cm}
\centering
\begin{overpic}[width=\figwidth]{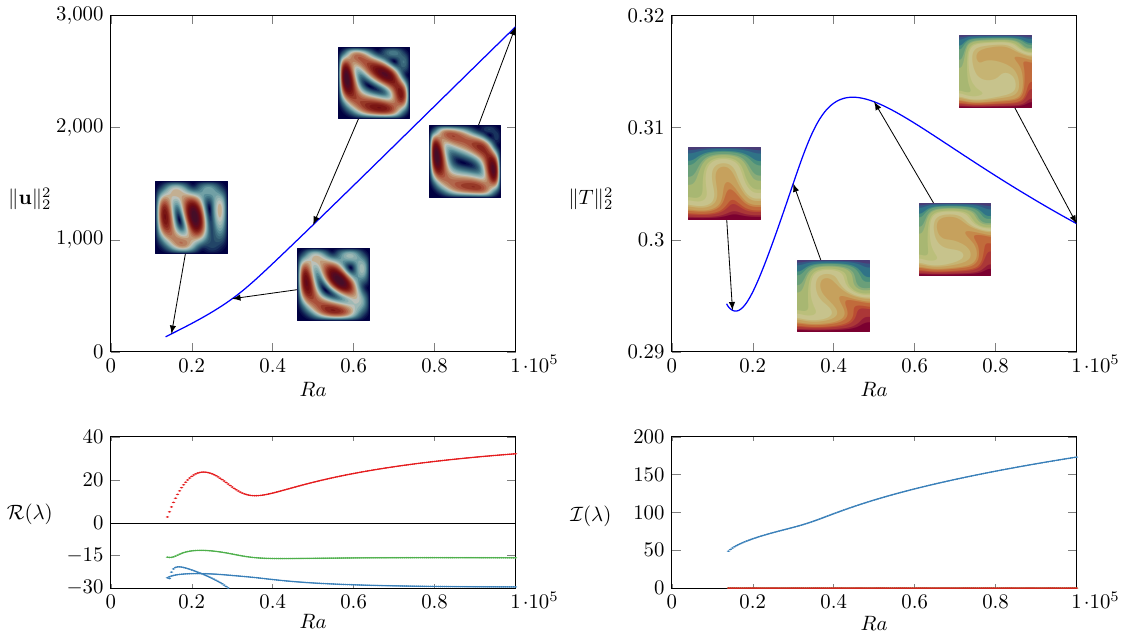}
\put(1,56){(a)}
\put(51,56){(b)}
\put(1,18){(c)}
\put(51,18){(d)}
\end{overpic}
\caption{(Color online) Evolution of the steady states in branch (10), bifurcating from branch (2) at $Ra \approx 13550$, illustrated via (a) the kinetic energy and (b) the potential energy of branch (10). The largest growth rates and corresponding frequencies are presented in (c) and (d), respectively.}
\label{fig:branch_8}
\end{figure}

The second eigenmode introduced in \cref{fig:eigenmodes} gives birth to branch (2) at $Ra_c^{(2)} = 6746$, which is depicted in \cref{fig:branch_0}(a)-(b). The bifurcation is transcritical and preserves the symmetries of the eigenmode in the velocity and temperature profiles. Then, the largest growth rate in \cref{fig:branch_0}(c) indicates that the solutions on this branch are stable over the interval $Ra \in [13500,85500]$ and unstable outside. A bifurcation occurs at $Ra \approx 13500$ when $\mathcal R(\lambda)= \mathcal I(\lambda) = 0$, which is analysed later and illustrated in \cref{fig:branch_8}. The bifurcation from branch (2) to branch (10) is accompanied by a loss of Boussinesq symmetry. Another bifurcation is observed at $Ra\approx 85500$ and has been obtained by deflated continuation. 
Figure \ref{fig:branch_8}(a) demonstrates that the two symmetries of the second eigenmode (see \cref{sec_problem}) are rapidly broken as $Ra$ increases after the bifurcation from branch (2) to the secondary branch (10). This symmetry breaking leads to a primary large scale circulation spanning the domain and a secondary vortex with smaller amplitude in one of the corners depending on which symmetric solution one is referring to. As $Ra \to 10^5$, the difference in the flow pattern of the velocity magnitude between branch (1) and (10) is the secondary vortex in the corner of the convection cell.
The symmetry breaking is also obvious in the temperature field (see \cref{fig:branch_8}(b)) with the potential energy of the flow increasing and then decreasing slightly in contrast to the increase of the kinetic energy and the convective heat transfer (see also \cref{fig:bif_diagrams}(c)). The largest growth rate shows that the states are unstable to a real eigenmode (see \cref{fig:branch_8}(d)). The subsequent growth rates shown in \cref{fig:branch_8}(c) are negative and almost constant in this range of Rayleigh numbers.

\begin{figure}[htbp]
\vspace{0.2cm}
\centering
\begin{overpic}[width=\figwidth]{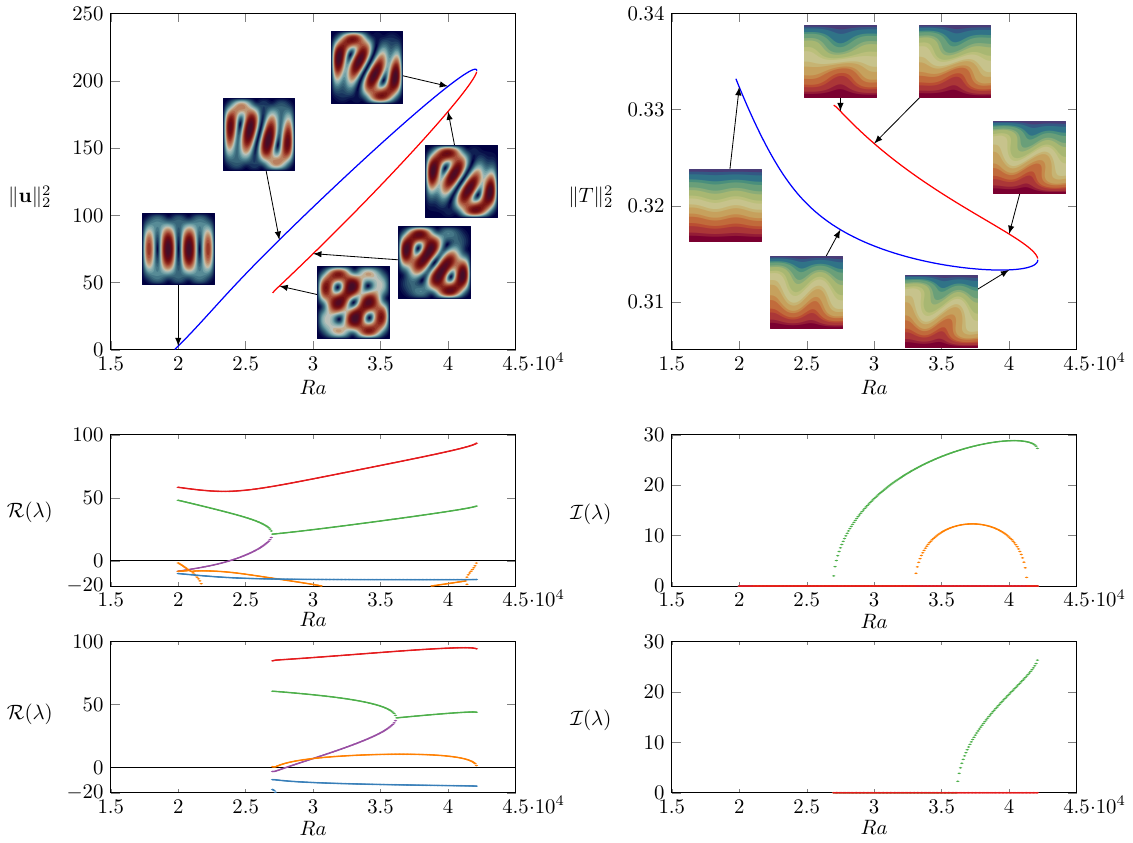}
\put(0,75){(a)}
\put(50,75){(b)}
\put(0,36){(c)}
\put(50,36){(d)}
\put(0,17){(e)}
\put(50,17){(f)}
\end{overpic}
\caption{(Color online) Evolution of the steady states in branch (3) arising from the third eigenmode, illustrated via (a) the kinetic energy and (b) the potential energy. The upper and lower parts of the branch are coloured in blue and red, respectively. The largest growth rates and corresponding frequencies are presented in (c) and (d), respectively.}
\label{fig:branch_turning_3}
\end{figure}

\begin{figure}[htbp]
\vspace{0.2cm}
\centering
\begin{overpic}[width=\figwidth]{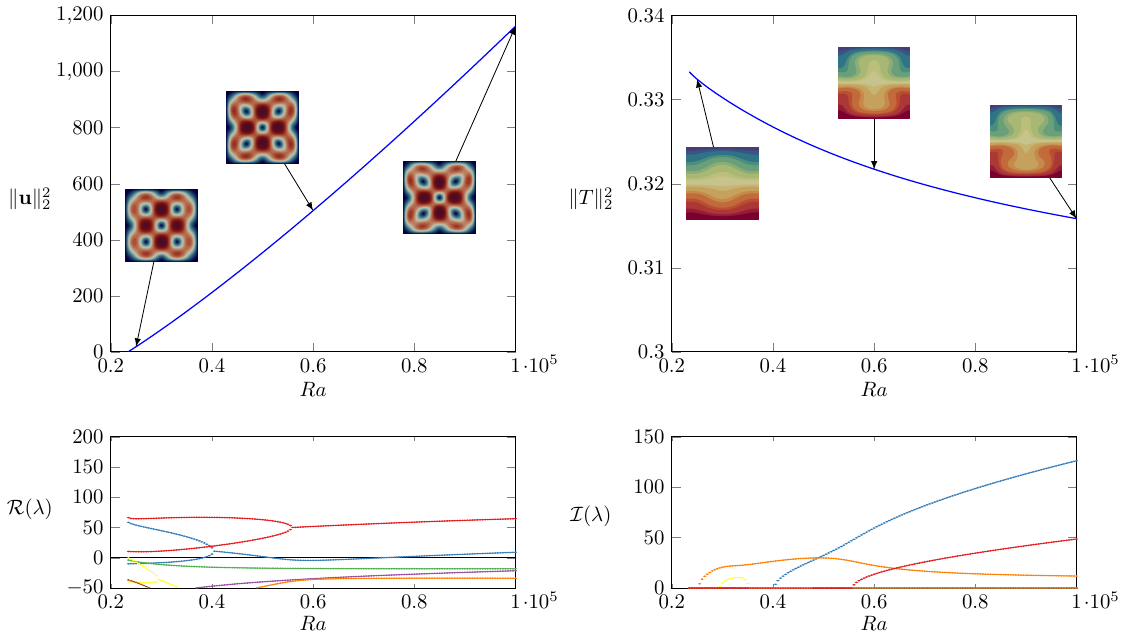}
\put(1,56){(a)}
\put(51,56){(b)}
\put(1,18){(c)}
\put(51,18){(d)}
\end{overpic}
\caption{(Color online) Evolution of the steady states in branch (4) arising from the fourth eigenmode, illustrated via (a) the kinetic energy and (b) the potential energy. The largest growth rates and corresponding frequencies are presented in (c) and (d), respectively.}
\label{fig:branch_d}

\vspace{1.2cm}
\centering
\begin{overpic}[width=\figwidth]{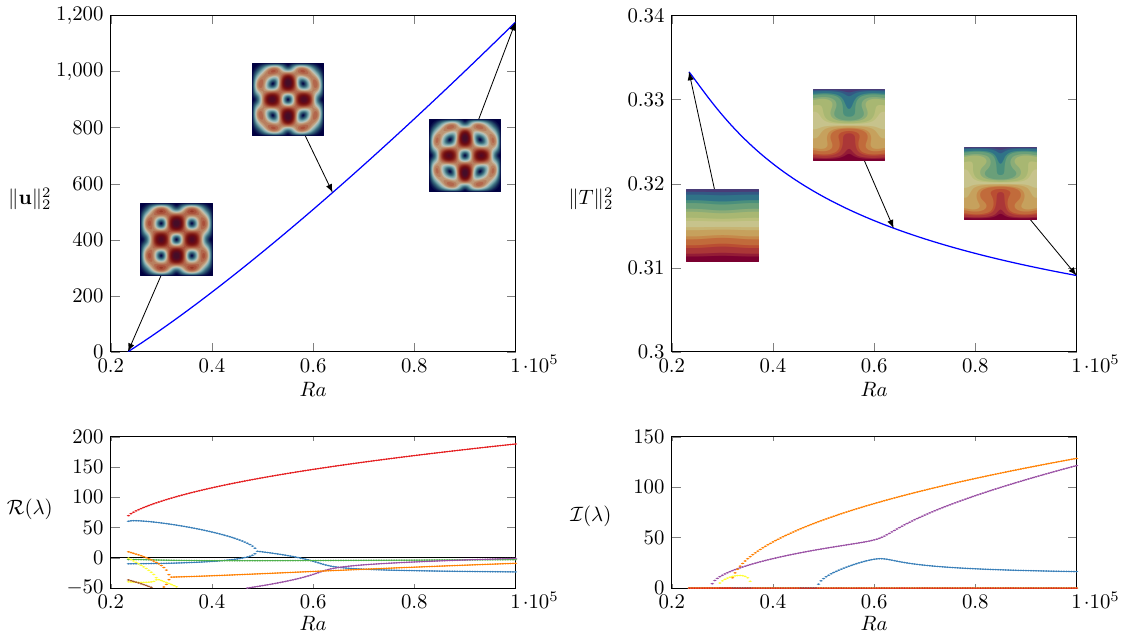}
\put(1,56){(a)}
\put(51,56){(b)}
\put(1,18){(c)}
\put(51,18){(d)}
\end{overpic}
\caption{(Color online) Evolution of the steady states in branch (5) illustrated via (a) the kinetic energy and (b) the potential energy. Similarly to branch (4), this branch bifurcates from the fourth eigenmode. The largest growth rates and corresponding frequencies are presented in (c) and (d), respectively.}
\label{fig:branch_12}
\end{figure}

Branch (3) originates from the eigenmode of the conducting state at $Ra_c^{(3)}=19655$, which is depicted in \cref{fig:branch_turning_3}. This branch preserves the centro-symmetry of the 3rd eigenmode and has a turning point located at $Ra\approx 42130$. Interestingly, the lower part of the branch, coloured in red in \cref{fig:branch_turning_3}(a,b) bifurcates from branch (5) around $Ra \approx 27000$. The latter branch emanates from the fourth eigenmode and is illustrated later in \cref{fig:branch_12}. The largest growth rates of the upper and lower parts of branch (3) are displayed in \cref{fig:branch_turning_3}(c,d), together with their associated frequencies. We find that the steady states in the branch are unstable with a zero frequency throughout the range of Rayleigh numbers for which the branch exists. Moreover, we observe that the eigenmode coloured in purple in \cref{fig:branch_turning_3}(c) traverses zero at $Ra\approx 23800$, indicating the presence of a secondary bifurcation, which we will analyse in the following paragraph. We point out that this branch has not been originally discovered by deflated continuation (our method does not guarantee to find all the steady states of a problem) but obtained by continuing the third linear state with finer steps in the Rayleigh number. This illustrates that deflated continuation may be complemented by standard bifurcation analysis techniques, such as arclength continuation and branch-switching algorithms.
\begin{figure}[htbp]
\vspace{0.2cm}
\centering
\begin{overpic}[width=\figwidth]{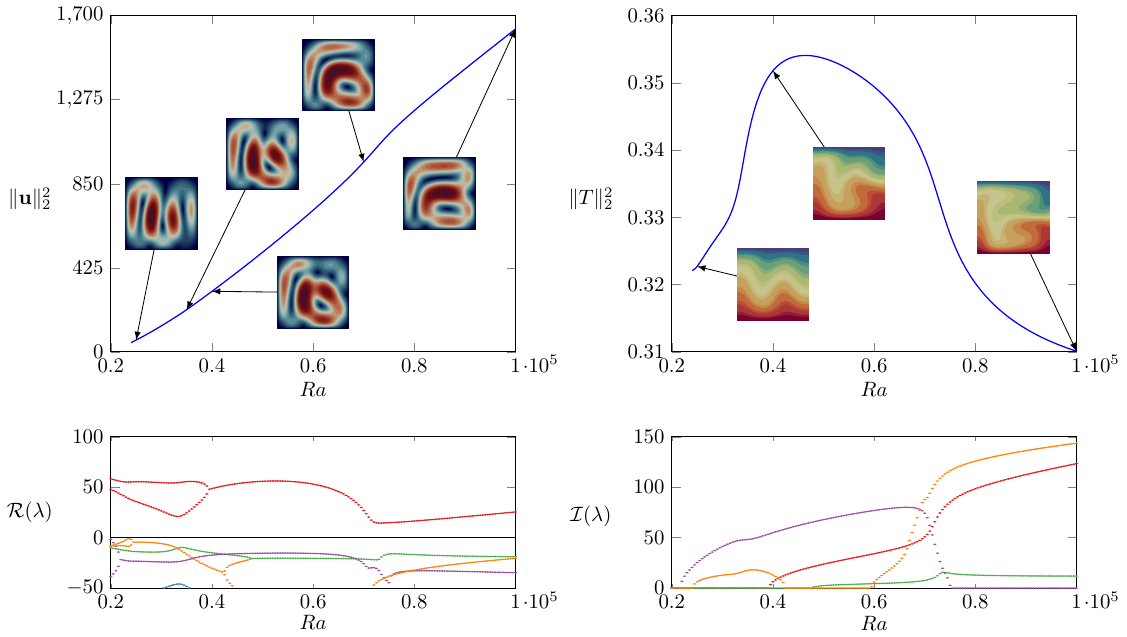}
\put(1,56){(a)}
\put(51,56){(b)}
\put(1,18){(c)}
\put(51,18){(d)}
\end{overpic}
\caption{(Color online) Evolution of the steady states in branch (12), bifurcating from branch (3) around $Ra\approx 23800$, and illustrated via (a) the kinetic energy and (b) the potential energy. The largest growth rates and corresponding frequencies are presented in (c) and (d), respectively.}
\label{fig:branch_c}
\end{figure}
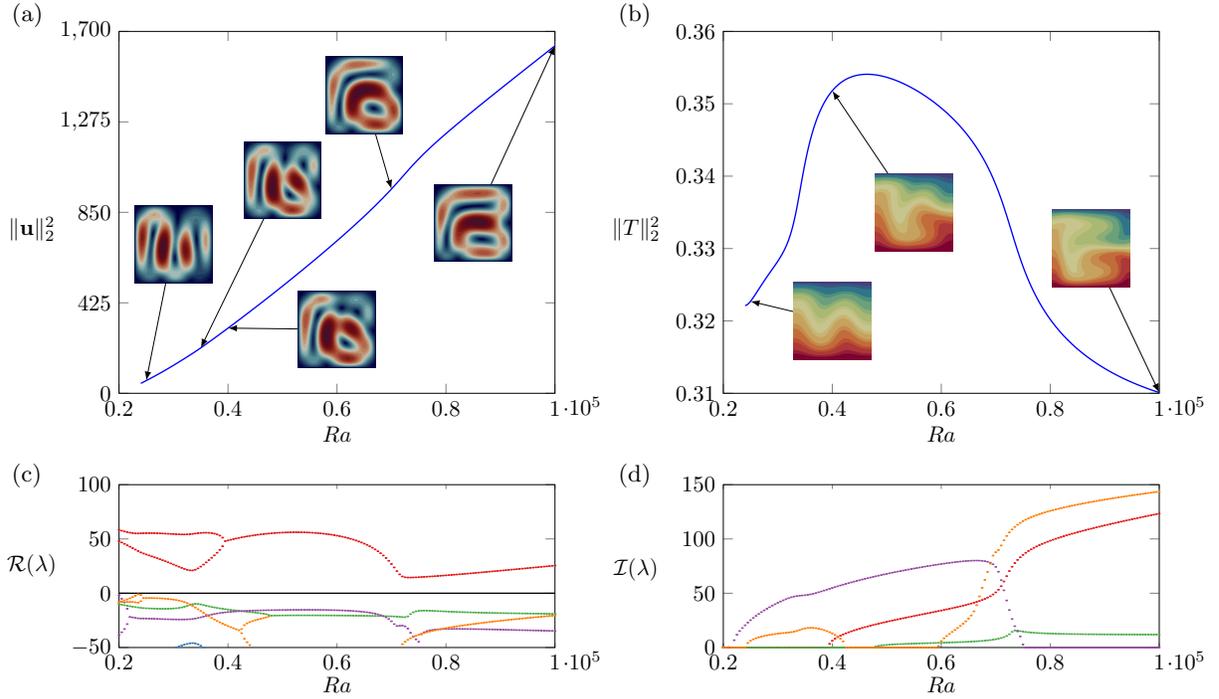
We now focus on branch (12), which bifurcates from branch (3) at $Ra\approx 23800$, and display the evolution of the steady states in the branch in \cref{fig:branch_c} along with its stability analysis. This branch breaks the centro-symmetry of branch (3) and the pattern of the velocity magnitude becomes more complex as $Ra$ increases. This complex pattern is one reason for the meandering of the potential energy in this range of Rayleigh numbers and enhances the convective heat transfer (see \cref{fig:bif_diagrams}(c)) by mixing the temperature field. \cref{fig:branch_c}(c) shows that the steady states of branch (12) are unstable to two eigenmodes until $Ra \approx 39500$, where they coalesce. At this Rayleigh number, we see that the two most unstable eigenmodes coalesce into an unstable complex conjugate pair of eigenvalues (see \cref{fig:branch_c}(d)). All the subsequent eigenmodes are found to be stable with either $\mathcal I(\lambda) = 0$ or $\mathcal I(\lambda) \neq 0$.

\cref{fig:branch_d} shows bifurcation diagrams and the linear stability results of branch (4) which originates from the fourth eigenmode at $Ra_c^{(4)} = 23346$ through a transcritical bifurcation. The flow pattern of the velocity magnitude has a symmetric form of an array of four vortices, which is conserved over the whole range of the Rayleigh numbers we consider as \cref{fig:branch_d}(a) suggests. Similarly, the symmetric pattern of the temperature field remains unaffected in the regime $23346 \leq Ra \leq 10^5$ (see \cref{fig:branch_d}(b)). We have verified numerically that the symmetries are conserved within this range of $Ra$. \cref{fig:branch_d}(c) shows the four largest growth rates as a function of $Ra$. At Rayleigh numbers close to $Ra_c^{(4)} = 23346$ we see that three of the eigenmodes are real and unstable, while the fourth one is real and stable. As $Ra$ increases, the two most unstable eigenmodes (coloured in red and green, respectively, in \cref{fig:branch_d}(c)) coalesce at $Ra \approx 56000$ into a complex unstable eigenmode whose growth rate and oscillation frequency increases as $Ra \to 10^5$. Similarly, the other two real unstable modes coalesce at $Ra \approx 40000$ into a complex conjugate pair of eigenmodes which are unstable for $40000 < Ra \leq 10^5$. At $Ra\approx 52000$ and $Ra\approx 73500$, we observe in \cref{fig:branch_d}(c) that the growth rate coloured in blue vanishes with a nonzero frequency, leading to Hopf bifurcations where periodic solutions to the time-dependent problem \eqref{eq_RB} arise.

\begin{figure}[htbp]
\vspace{0.2cm}
\centering
\begin{overpic}[width=\figwidth]{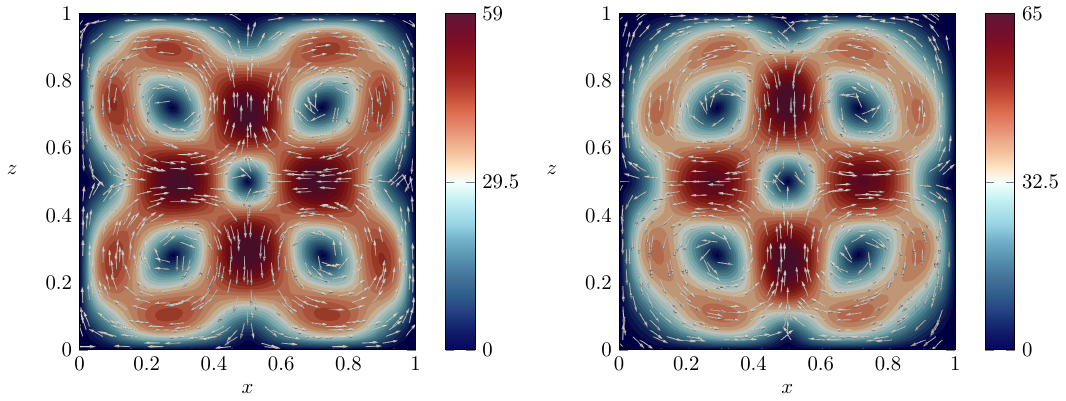}
\put(19,38){Branch (4)}
\put(69,38){Branch (5)}
\end{overpic}
\caption{(Color online) Velocity magnitude of the steady state in branches (4) and (5) at $Ra=10^5$.}
\label{fig_comparison_velocity}
\end{figure}

We close our discussion on the branches arising from the conducting steady state with branch (5), presented in \cref{fig:branch_12}. This branch originates from the fourth eigenmode of the conducting state at $Ra \approx 23346$ through the same transcritical bifurcation as branch (4). The steady states in branch (5) conserve the Boussinesq and $x$ symmetries, with a velocity field consisting of four vortices, located symmetrically at each corner of the domain. The velocity fields of branches (4) and (5), illustrated in \cref{fig_comparison_velocity}, are different, indicating that branch (5) is not a symmetric version of branch (4). The stability analysis reveals that the steady states in this branch are very sensitive to perturbations and unstable to a real eigenmode throughout the range of $Ra$ considered in this study. Moreover, the largest growth rate at $Ra=10^5$ is equal to $\mathcal{R}(\lambda) \approx 188$, which is approximately three times larger than the largest growth rate of the steady state in branch (4) at the same Rayleigh number (see \cref{fig:branch_d}(c) and \cref{fig:branch_12}(c)). One of the eigenmodes vanishes at $Ra\approx 27000$ giving rise to branch (3), depicted in \cref{fig:branch_turning_3}. Finally, the real part of the eigenvalue depicted in blue in \cref{fig:branch_12}(c-d) crosses zero at $Ra\approx 57500$, which shows the existence of a Hopf bifurcation.

\begin{figure}[htbp]
\vspace{0.2cm}
\centering
\begin{overpic}[width=\figwidth]{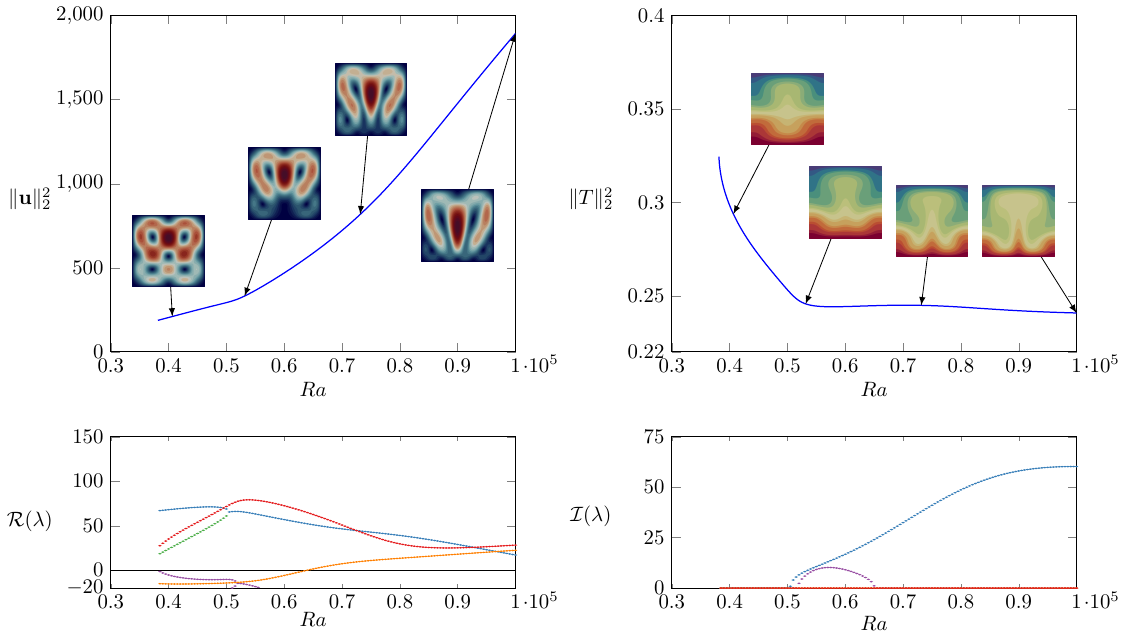}
\put(1,56){(a)}
\put(51,56){(b)}
\put(1,18){(c)}
\put(51,18){(d)}
\end{overpic}
\caption{(Color online) Evolution of the steady states in branch (6), bifurcating from branch (4) at $Ra \approx 38150$, illustrated via (a) the kinetic energy and (b) the potential energy. The largest growth rates and corresponding frequencies are presented in (c) and (d), respectively.}
\label{fig:branch_13}
\vspace{1.2cm}
\centering
\begin{overpic}[width=\figwidth]{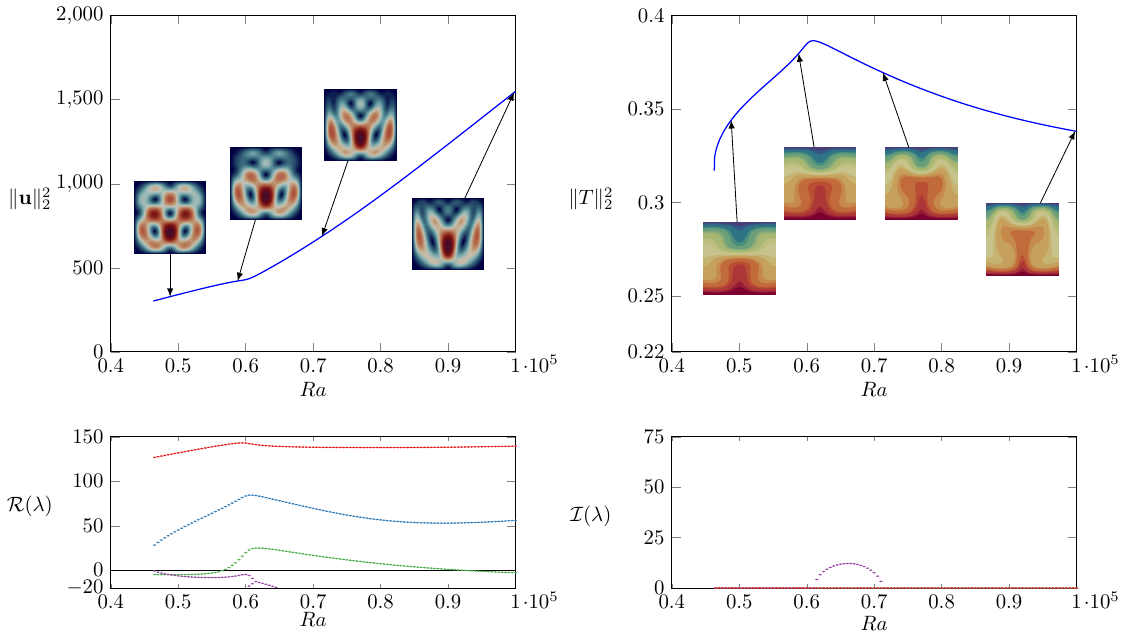}
\put(1,56){(a)}
\put(51,56){(b)}
\put(1,18){(c)}
\put(51,18){(d)}
\end{overpic}
\caption{(Color online) Evolution of the steady states in branch (7), bifurcating from branch (5) at $Ra \approx 46300$, illustrated via (a) the kinetic energy and (b) the potential energy. The largest growth rates and corresponding frequencies are presented in (c) and (d), respectively.}
\label{fig:branch_14}
\end{figure}

We now analyze the branches bifurcating from branches (4) and (5). The fourth largest growth rate of branch (4), coloured in blue in \cref{fig:branch_d}(c), crosses zero at $Ra \approx 38150$, leading to the bifurcating branch (6) depicted in \cref{fig:branch_13}. We observe that the Boussinesq symmetry of branch (4) is broken by the bifurcation, while the symmetry with respect to the $x$ axis is preserved. Additionally, the linear stability analysis indicates that this branch is unstable with a zero frequency except in the interval $Ra\in [72500,93000]$, where the largest growth rate is associated with a positive frequency, as shown by the eigenvalues coloured in blue in \cref{fig:branch_13}(a) and (c). Moreover, the third largest growth rate (coloured in orange) crosses zero around $Ra\approx 63500$, indicating the presence of a bifurcation which has not been discovered by deflation.
Finally, a branch, denoted by branch (7), bifurcates from branch (5) at $Ra \approx 46300$. The steady states in this branch are illustrated in \cref{fig:branch_14}. We observe on the velocity magnitude and temperature field that the symmetry of branch (5) with respect to the $x$ axis is maintained while the Boussinesq symmetry is broken. Branches (6) and (7) might seem nearly symmetric but they arise at different Rayleigh numbers and have different eigenvalues. The linear stability analysis of branch (7) shows that the steady states are unstable to a real eigenmode, with a largest growth rate around $\mathcal{R}(\lambda)\approx 150$ throughout the range of $Ra$ considered in this study. In \cref{fig:branch_14}(c) and (d), we observe that the third eigenvalue (plotted in green) crosses zero at $Ra\approx 92000$, resulting in a tertiary branch that breaks the remaining symmetry. The branch has been discovered by deflation but we chose to not report it in the study.

\subsection{Disconnected branches}

\begin{figure}[htbp]
\vspace{0.2cm}
\centering
\begin{overpic}[width=\figwidth]{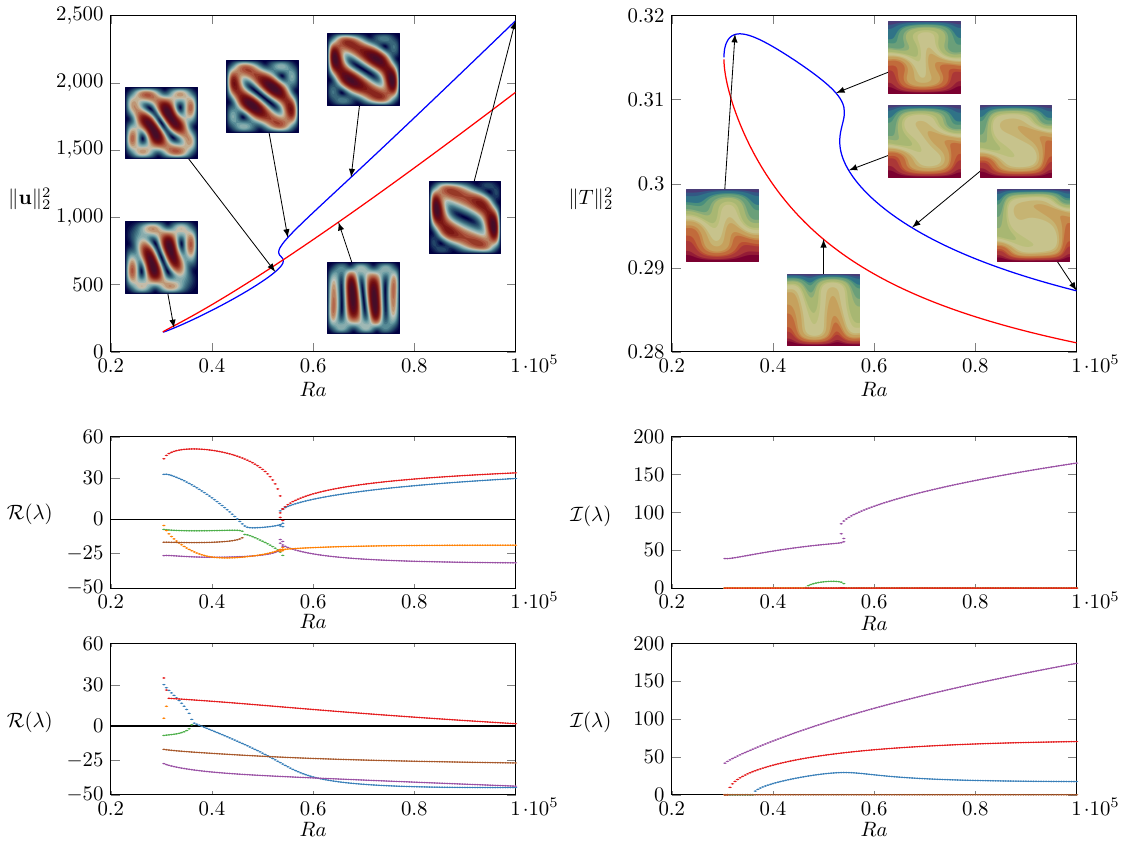}
\put(1,75){(a)}
\put(51,75){(b)}
\put(1,36){(c)}
\put(51,36){(d)}
\put(1,17){(e)}
\put(51,17){(f)}
\end{overpic}
\caption{(Color online) Evolution of the steady states in the disconnected branch (11), illustrated via (a) the kinetic energy and (b) the potential energy. The upper and lower parts of the branch are coloured in blue and red, respectively. The largest growth rates and corresponding frequencies of the upper (resp.~lower) part of the branch are presented in (c) and (d) (resp. (e) and (f)), respectively.}
\label{fig:branch_f}
\end{figure}

We now focus on the disconnected branch (11) (see \cref{fig:bif_diagrams}), which comprises an upper and a lower branch, depicted in blue and red in \cref{fig:branch_f}, respectively. At the upper (blue) branch the flow pattern of the velocity magnitude is a three vortex state with one large titled vortex spanning the domain and two smaller vortices located at the corners. As $Ra \to 10^5$ we notice that the state looks similar to the branches (1) and (10). The difference is that branch (10) involves only one smaller vortex at the corner of the domain and branch (1) has no vortices at the corners, while branch (11) has two smaller vortices located at the corners. At $Ra \approx 52000$ the branch takes the form of an S-shaped curve with hysteresis, similar to the cusp bifurcation~\cite{kuznetsov13}.

As the saddle-node bifurcation is approached from below, the growth rate of the most unstable eigenmode decreases toward zero at the critical point (see \cref{fig:branch_f}(c)). Above the bifurcation point $\mathcal R(\lambda)$ increases and seems to saturate as $Ra \to 10^5$. The frequency corresponding to the growth rate is zero in the whole range of $Ra$ numbers suggesting that the steady state is unstable to a real eigenmode in this regime. The potential energy of the upper (blue) branch decreases with $Ra$ in contrast to the kinetic energy. As $Ra$ increases the flow pattern of the temperature becomes more efficient at transferring heat due to the large scale circulation in the flow (see \cref{fig:bif_diagrams}(c)). The flow pattern of the velocity of the upper (blue) branch, shown for $Ra = 10^5$ (see \cref{fig:branch_f}(a)), exhibits a large scale vortex and two smaller vortices at the corners of the domain. This is reminiscent of the flow patterns from studies on the dynamics of random-in-time flow reversals \cite{kadanoff01,sugiyamaetal10,chandraverma13,podvinsergent17}. 
These reversals in the flow direction of the large scale circulation at irregular time intervals have been observed in domains with free-slip or no-slip boundaries and in a narrow band of Rayleigh numbers in the range $10^7 \leq Ra \leq10^9$ depending on the Prandtl number \cite{sugiyamaetal10,chandraverma13,podvinsergent17}. In this regime the system fluctuates in time between essentially the state we show for $Ra = 10^5$ (see \cref{fig:branch_f}(a)) and its mirror symmetric state. These large changes in the dynamics of the flow are caused by bifurcations over a turbulent background \cite{fauveetal17}. Such bifurcations have been observed in many other types of flows \cite{raveletetal04,berhanuetal07,cadotetal15,dallasetal19,wdp20} and understanding their dynamics is an open question of great interest.

At the lower (red) branch illustrated in \cref{fig:branch_f}(a) we observe the central vortex to be oriented more vertically, giving space to the smaller vortices at the corners to extend along the $z$ axis and almost reach the height of the domain. This flow pattern is an extension of the two-vortex pattern observed in branch (5) (see \cref{fig:branch_8}) and it has similar features to the three-vortex pattern found near the linear limit of branch (3) (see \cref{fig:branch_turning_3}). The patterns of the kinetic and potential energy of this branch are almost invariant within the range $30000 \lesssim Ra \leq 10^5$. The flow pattern of the temperature clearly manages to effectively push the hot fluid toward the cold plate at the top and vice versa. This efficiency in convective heat transfer is clearly depicted in \cref{fig:bif_diagrams}(c), where this part of branch (11) has the highest Nusselt number for $Ra > 95000$. Analysing the stability of the lower (red) branch, we find that the steady solutions become increasingly more stable as the Rayleigh number increases, with $\mathcal R(\lambda) \to 0$ at $Ra \approx 10^5$, leading to a Hopf bifurcation (see \cref{fig:branch_f}(e)). On the other hand, the frequency corresponding to the growth rate satisfies $\mathcal I(\lambda) = 0$ at the threshold of the disconnected branch and as $Ra$ increases the instability becomes progressively oscillatory (see \cref{fig:branch_f}(f)). The subsequent growth rates presented in \cref{fig:branch_f}(e) remain negative as $Ra$ increases suggesting that this part of branch (11) is another preferred steady state of the system particularly for $Ra > 10^5$. 

\begin{figure}[htbp]
\vspace{0.2cm}
\centering
\begin{overpic}[width=\figwidth]{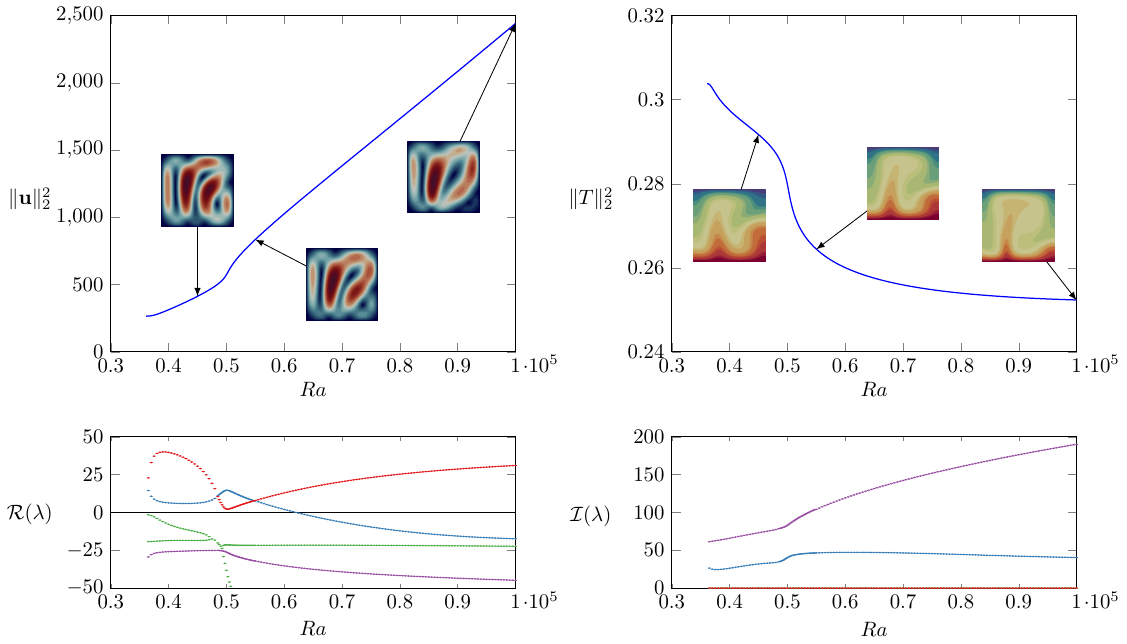}
\put(1,56){(a)}
\put(51,56){(b)}
\put(1,18){(c)}
\put(51,18){(d)}
\end{overpic}
\caption{(Color online) Evolution of the steady states in branch (8), bifurcating from the lower part of the disconnected branch (11) at $Ra \approx 37000$, illustrated via (a) the kinetic energy and (b) the potential energy. The largest growth rates and corresponding frequencies are presented in (c) and (d), respectively.}
\label{fig:branch_g}
\end{figure}

Branch (8) illustrated in \cref{fig:branch_g} bifurcates from the lower part of the disconnected branch (11) at $Ra \approx 37000$ (see also \cref{fig:bif_diagrams}). In this case, the pattern of the velocity transitions into a main bent vortex with a vertically elongated vortex on its left and a smaller vortex at the bottom right corner (see \cref{fig:branch_g}(a)). From \cref{fig:bif_diagrams} we can see that even though the kinetic energy of branch (8) is equal to that of the upper part of branch (11) for $Ra > 50000$, the potential energy is much smaller and apparently the smallest among the branches we have considered. The pattern of the temperature is essentially a hot plume that rises effectively all the way to the cold plate (see \cref{fig:branch_g}(b)). This flow pattern becomes increasingly more efficient at transferring heat across the domain as $Ra$ increases (see \cref{fig:bif_diagrams}(c)). However, it is not as efficient in convecting heat as the pattern of branch (11) (see lower branch in \ref{fig:branch_f}(b)) because two hot plumes are obviously more effective than one. The steady state of branch (8) is unstable to a real eigenmode over the range $37000 \lesssim Ra \leq 10^5$ except for a small range around $Ra = 50000$ where it becomes unstable to a complex eigenmode. This can be seen from \cref{fig:branch_g}(c) and \ref{fig:branch_g}(d), where the two largest growth rates (coloured in red and blue) swap with each other in this small range of Rayleigh numbers. Finally, the real part of the eigenvalue depicted in blue in \cref{fig:branch_g}(c-d) crosses zero at $Ra\approx 62000$, giving birth to a Hopf bifurcation.

\begin{figure}[htbp]
\vspace{0.2cm}
\centering
\begin{overpic}[width=\figwidth]{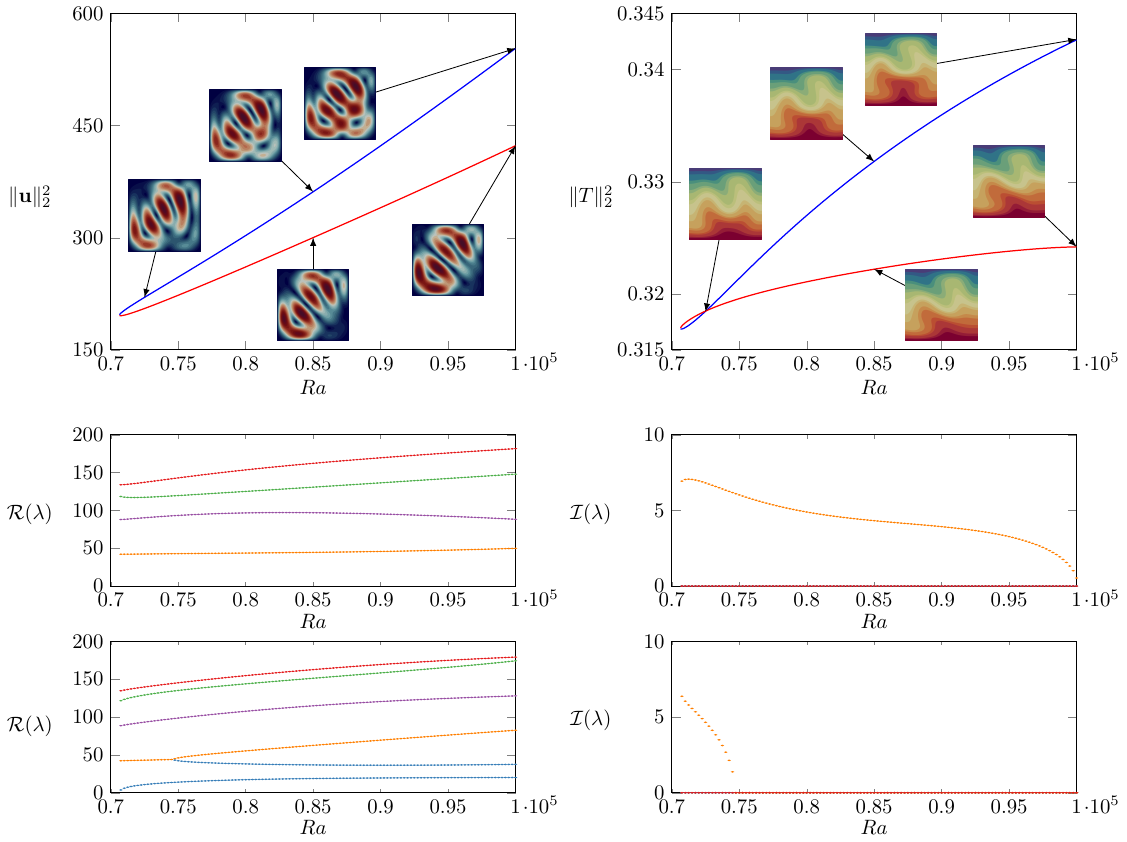}
\put(1,75){(a)}
\put(51,75){(b)}
\put(1,36){(c)}
\put(51,36){(d)}
\put(1,17){(e)}
\put(51,17){(f)}
\end{overpic}
\caption{(Color online) Evolution of the steady states in the disconnected branch (9), illustrated via (a) the kinetic energy and (b) the potential energy. The upper and lower parts of the branch are coloured in blue and red, respectively. The largest growth rates and corresponding frequencies of the upper (resp.~lower) part of the branch are presented in (c) and (d) (resp. (e) and (f)), respectively.}
\label{fig:branch_h}
\end{figure}

The second disconnected branch that we find in this range of Rayleigh numbers is branch (9) and arises at $Ra \approx 70000$ (see \cref{fig:branch_h}). From \cref{fig:branch_h}(a), we see that all the symmetries of the flow are broken in this branch and the flow pattern consists of a number of vortices with different orientations. These flow structures are not persistent over the range of parameters that we consider, and hence the heat is not transferred effectively across the domain (see also \cref{fig:bif_diagrams}(c)). Here, \cref{fig:branch_h} shows that the potential energy of branch (9) increases along with the kinetic energy unlike in most of the other branches we considered and from \cref{fig:bif_diagrams}(b) we see that it is significant in comparison to many other branches. Moreover, from the linear stability analysis of the steady states we find that both the upper and lower parts of this disconnected branch are unstable to a real eigenmode, i.e. $\mathcal R(\lambda) > 0$ and $\mathcal I(\lambda) = 0$, with large growth rates. This suggests that this branch is very unlikely to be observed in a laboratory experiment, and highlights the effectiveness of deflated continuation in identifying such unstable disconnected branches and steady states.

\section{Conclusions} \label{sec_conclusions}

In this work, we have computed and analysed the linear stability of steady states of the two-dimensional Rayleigh--B\'enard convection in a square cell with rigid walls. The combination of deflated continuation and a suitable initialization procedure based on the linear stability of the motionless state is revealed to be a powerful numerical technique for finding multiple solutions including disconnected branches, which are not easily accessible by standard bifurcation techniques such as arclength continuation and time-evolution algorithms. A highlight is the success of deflation to discover these branches in an automated manner without any prior knowledge of the dynamics.

We classify the solutions based on the kinetic and potential energy as well as the Nusselt number. Our classification provides a clear view of the steady states of Rayleigh--B\'enard convection up to a Rayleigh number of $10^5$. We show that the steady states of branch (1) (which originates from the conducting state) dictate the dynamics of the flow for $Ra \leq 10^5$, while for $Ra \sim 10^5$ the flow patterns of the disconnected branch (11) start to play a very important role as the steady states become stable. Solutions on branch (11) are reminiscent of the flow pattern that plays the fundamental role in the turbulent dynamics of flow reversals that have been reported in the literature for Rayleigh--B\'enard convection at $Ra \sim 10^7$~\cite{sugiyamaetal10,chandraverma13,podvinsergent17}. This disconnected branch exhibits an S-shaped curve with hysteresis which can be a challenge to discover with other bifurcation techniques.

There are several possible extensions to this work. An interesting but computationally challenging future study is to extend certain solutions of interest to higher values of $Ra$ and determine whether they give the required theoretical scalings in the limit of high Rayleigh number. A first attempt toward this direction has been done in Rayleigh--B\'enard convection with periodic boundary conditions in the horizontal direction and no-slip in the vertical direction \cite{sondaketak15,waleffeetal15}. Moreover, the extensions of branch (11) and its stability in the regime where reversals occur is also of great interest.
Another extension of interest is to study how the states presented in this study change with respect to other parameters of the system such as the Prandtl number or the aspect ratio of the domain $\Gamma$. As observed in the left panel of \cref{fig:critical_ratio}, the critical Rayleigh number converges to $Ra_c^*\approx 1707.762$ for large aspect ratio. Using alternative formulations of the Rayleigh--B\'enard problem, with the aspect ratio playing the role of the bifurcation parameter, it is possible to analyse the evolution of the branches for large aspect ratio and connect them with the solutions of the periodic set-up, which has been studied well in the literature, both analytically~\cite{chandrasekhar1961hydrodynamic} and numerically~\cite{zienicke1998bifurcations,paul2012bifurcation}. This connection will be of great interest to theoreticians and hopefully will spark their interest to explore analytical solutions on more realistic convection set-ups with side walls.

A different future direction would be to perform bifurcation analysis on Rayleigh--B\'enard convection in a three-dimensional rectangular domain. Such an approach will be of interest to experimentalists since direct comparisons with experiments would be possible. However, while the generalisation of the bifurcation technique employed in this paper to three dimensions is straightforward, high Rayleigh numbers require an efficient solver to perform the underlying Newton iterations. One possible solution would be to construct an efficient preconditioner robust with respect to the Rayleigh number to be able to reach high $Ra$ regimes in the spirit of~\cite{farrell2020augmented,farrell2019augmented}.

\begin{acknowledgments}
We would like to thank P.~A.~Gazca-Orozco and S.~Fauve for helpful discussions that improved the quality of the manuscript. We thank the anonymous referee for the detailed suggestions that greatly improved the quality of our manuscript. This work is supported by the EPSRC Centre For Doctoral Training in Industrially Focused Mathematical Modelling (EP/L015803/1) in collaboration with Simula Research Laboratory, and by EPSRC grants EP/V001493/1 and EP/R029423/1.
\end{acknowledgments}

\bibliography{biblio}

\begin{thebibliography}{56}%
\makeatletter
\providecommand \@ifxundefined [1]{%
 \@ifx{#1\undefined}
}%
\providecommand \@ifnum [1]{%
 \ifnum #1\expandafter \@firstoftwo
 \else \expandafter \@secondoftwo
 \fi
}%
\providecommand \@ifx [1]{%
 \ifx #1\expandafter \@firstoftwo
 \else \expandafter \@secondoftwo
 \fi
}%
\providecommand \natexlab [1]{#1}%
\providecommand \enquote  [1]{``#1''}%
\providecommand \bibnamefont  [1]{#1}%
\providecommand \bibfnamefont [1]{#1}%
\providecommand \citenamefont [1]{#1}%
\providecommand \href@noop [0]{\@secondoftwo}%
\providecommand \href [0]{\begingroup \@sanitize@url \@href}%
\providecommand \@href[1]{\@@startlink{#1}\@@href}%
\providecommand \@@href[1]{\endgroup#1\@@endlink}%
\providecommand \@sanitize@url [0]{\catcode `\\12\catcode `\$12\catcode
  `\&12\catcode `\#12\catcode `\^12\catcode `\_12\catcode `\%12\relax}%
\providecommand \@@startlink[1]{}%
\providecommand \@@endlink[0]{}%
\providecommand \url  [0]{\begingroup\@sanitize@url \@url }%
\providecommand \@url [1]{\endgroup\@href {#1}{\urlprefix }}%
\providecommand \urlprefix  [0]{URL }%
\providecommand \Eprint [0]{\href }%
\providecommand \doibase [0]{https://doi.org/}%
\providecommand \selectlanguage [0]{\@gobble}%
\providecommand \bibinfo  [0]{\@secondoftwo}%
\providecommand \bibfield  [0]{\@secondoftwo}%
\providecommand \translation [1]{[#1]}%
\providecommand \BibitemOpen [0]{}%
\providecommand \bibitemStop [0]{}%
\providecommand \bibitemNoStop [0]{.\EOS\space}%
\providecommand \EOS [0]{\spacefactor3000\relax}%
\providecommand \BibitemShut  [1]{\csname bibitem#1\endcsname}%
\let\auto@bib@innerbib\@empty
%</preamble>
\bibitem [{\citenamefont {Cross}\ and\ \citenamefont
  {Hohenberg}(1993)}]{crosshohenberg93}%
  \BibitemOpen
  \bibfield  {author} {\bibinfo {author} {\bibfnamefont {M.~C.}\ \bibnamefont
  {Cross}}\ and\ \bibinfo {author} {\bibfnamefont {P.~C.}\ \bibnamefont
  {Hohenberg}},\ }\bibfield  {title} {\bibinfo {title} {Pattern formation
  outside of equilibrium},\ }\href
  {https://link.aps.org/doi/10.1103/RevModPhys.65.851} {\bibfield  {journal}
  {\bibinfo  {journal} {Rev. Mod. Phys.}\ }\textbf {\bibinfo {volume} {65}},\
  \bibinfo {pages} {851} (\bibinfo {year} {1993})}\BibitemShut {NoStop}%
\bibitem [{\citenamefont {Bodenschatz}\ \emph {et~al.}(2000)\citenamefont
  {Bodenschatz}, \citenamefont {Pesch},\ and\ \citenamefont
  {Ahlers}}]{bodenschatzetal00}%
  \BibitemOpen
  \bibfield  {author} {\bibinfo {author} {\bibfnamefont {E.}~\bibnamefont
  {Bodenschatz}}, \bibinfo {author} {\bibfnamefont {W.}~\bibnamefont {Pesch}},\
  and\ \bibinfo {author} {\bibfnamefont {G.}~\bibnamefont {Ahlers}},\
  }\bibfield  {title} {\bibinfo {title} {{Recent developments in
  Rayleigh-B{\'e}nard convection}},\ }\href
  {https://doi.org/10.1146/annurev.fluid.32.1.709} {\bibfield  {journal}
  {\bibinfo  {journal} {Annu. Rev. Fluid Mech.}\ }\textbf {\bibinfo {volume}
  {32}},\ \bibinfo {pages} {709} (\bibinfo {year} {2000})}\BibitemShut
  {NoStop}%
\bibitem [{\citenamefont {Ouertatani}\ \emph {et~al.}(2008)\citenamefont
  {Ouertatani}, \citenamefont {Cheikh}, \citenamefont {Beya},\ and\
  \citenamefont {Lili}}]{ouertatani2008numerical}%
  \BibitemOpen
  \bibfield  {author} {\bibinfo {author} {\bibfnamefont {N.}~\bibnamefont
  {Ouertatani}}, \bibinfo {author} {\bibfnamefont {N.~B.}\ \bibnamefont
  {Cheikh}}, \bibinfo {author} {\bibfnamefont {B.~B.}\ \bibnamefont {Beya}},\
  and\ \bibinfo {author} {\bibfnamefont {T.}~\bibnamefont {Lili}},\ }\bibfield
  {title} {\bibinfo {title} {{Numerical simulation of two-dimensional
  Rayleigh--B{\'e}nard convection in an enclosure}},\ }\href
  {https://doi.org/10.1016/j.crme.2008.02.004} {\bibfield  {journal} {\bibinfo
  {journal} {C. R. Mec.}\ }\textbf {\bibinfo {volume} {336}},\ \bibinfo {pages}
  {464} (\bibinfo {year} {2008})}\BibitemShut {NoStop}%
\bibitem [{\citenamefont {Zienicke}\ \emph {et~al.}(1998)\citenamefont
  {Zienicke}, \citenamefont {Seehafer},\ and\ \citenamefont
  {Feudel}}]{zienicke1998bifurcations}%
  \BibitemOpen
  \bibfield  {author} {\bibinfo {author} {\bibfnamefont {E.}~\bibnamefont
  {Zienicke}}, \bibinfo {author} {\bibfnamefont {N.}~\bibnamefont {Seehafer}},\
  and\ \bibinfo {author} {\bibfnamefont {F.}~\bibnamefont {Feudel}},\
  }\bibfield  {title} {\bibinfo {title} {{Bifurcations in two-dimensional
  Rayleigh-B{\'e}nard convection}},\ }\href
  {https://link.aps.org/doi/10.1103/PhysRevE.57.428} {\bibfield  {journal}
  {\bibinfo  {journal} {Phys. Rev. E}\ }\textbf {\bibinfo {volume} {57}},\
  \bibinfo {pages} {428} (\bibinfo {year} {1998})}\BibitemShut {NoStop}%
\bibitem [{\citenamefont {Paul}\ \emph {et~al.}(2012)\citenamefont {Paul},
  \citenamefont {Verma}, \citenamefont {Wahi}, \citenamefont {Reddy},\ and\
  \citenamefont {Kumar}}]{paul2012bifurcation}%
  \BibitemOpen
  \bibfield  {author} {\bibinfo {author} {\bibfnamefont {S.}~\bibnamefont
  {Paul}}, \bibinfo {author} {\bibfnamefont {M.~K.}\ \bibnamefont {Verma}},
  \bibinfo {author} {\bibfnamefont {P.}~\bibnamefont {Wahi}}, \bibinfo {author}
  {\bibfnamefont {S.~K.}\ \bibnamefont {Reddy}},\ and\ \bibinfo {author}
  {\bibfnamefont {K.}~\bibnamefont {Kumar}},\ }\bibfield  {title} {\bibinfo
  {title} {{Bifurcation analysis of the flow patterns in two-dimensional
  Rayleigh--B{\'e}nard convection}},\ }\href
  {https://doi.org/10.1142/S0218127412300182} {\bibfield  {journal} {\bibinfo
  {journal} {Int. J. Bifurcat. Chaos}\ }\textbf {\bibinfo {volume} {22}},\
  \bibinfo {pages} {1230018} (\bibinfo {year} {2012})}\BibitemShut {NoStop}%
\bibitem [{\citenamefont {Mishra}\ \emph {et~al.}(2010)\citenamefont {Mishra},
  \citenamefont {Wahi},\ and\ \citenamefont {Verma}}]{mishra2010patterns}%
  \BibitemOpen
  \bibfield  {author} {\bibinfo {author} {\bibfnamefont {P.~K.}\ \bibnamefont
  {Mishra}}, \bibinfo {author} {\bibfnamefont {P.}~\bibnamefont {Wahi}},\ and\
  \bibinfo {author} {\bibfnamefont {M.~K.}\ \bibnamefont {Verma}},\ }\bibfield
  {title} {\bibinfo {title} {{Patterns and bifurcations in low--Prandtl-number
  Rayleigh-B{\'e}nard convection}},\ }\href
  {https://doi.org/10.1209%2F0295-5075%2F89%2F44003} {\bibfield  {journal}
  {\bibinfo  {journal} {EPL}\ }\textbf {\bibinfo {volume} {89}},\ \bibinfo
  {pages} {44003} (\bibinfo {year} {2010})}\BibitemShut {NoStop}%
\bibitem [{\citenamefont {Peterson}(2008)}]{Peterson2008}%
  \BibitemOpen
  \bibfield  {author} {\bibinfo {author} {\bibfnamefont {J.~W.}\ \bibnamefont
  {Peterson}},\ }\emph {\bibinfo {title} {Parallel adaptive finite element
  methods for problems in natural convection}},\ \href
  {http://hdl.handle.net/2152/18091} {Ph.D. thesis},\ \bibinfo  {school}
  {University of Texas}, \bibinfo {address} {Austin} (\bibinfo {year}
  {2008})\BibitemShut {NoStop}%
\bibitem [{\citenamefont {Keller}(1977)}]{keller1977numerical}%
  \BibitemOpen
  \bibfield  {author} {\bibinfo {author} {\bibfnamefont {H.~B.}\ \bibnamefont
  {Keller}},\ }\bibfield  {title} {\bibinfo {title} {Numerical solution of
  bifurcation and nonlinear eigenvalue problems.},\ }in\ \href@noop {} {\emph
  {\bibinfo {booktitle} {Applications of Bifurcation Theory}}}\ (\bibinfo
  {publisher} {Academic Press},\ \bibinfo {year} {1977})\ pp.\ \bibinfo {pages}
  {359--384}\BibitemShut {NoStop}%
\bibitem [{\citenamefont {Ma}\ \emph {et~al.}(2006)\citenamefont {Ma},
  \citenamefont {Sun},\ and\ \citenamefont {Yin}}]{ma2006multiplicity}%
  \BibitemOpen
  \bibfield  {author} {\bibinfo {author} {\bibfnamefont {D.-J.}\ \bibnamefont
  {Ma}}, \bibinfo {author} {\bibfnamefont {D.-J.}\ \bibnamefont {Sun}},\ and\
  \bibinfo {author} {\bibfnamefont {X.-Y.}\ \bibnamefont {Yin}},\ }\bibfield
  {title} {\bibinfo {title} {{Multiplicity of steady states in cylindrical
  Rayleigh-B{\'e}nard convection}},\ }\href
  {https://link.aps.org/doi/10.1103/PhysRevE.74.037302} {\bibfield  {journal}
  {\bibinfo  {journal} {Phys. Rev. E}\ }\textbf {\bibinfo {volume} {74}},\
  \bibinfo {pages} {037302} (\bibinfo {year} {2006})}\BibitemShut {NoStop}%
\bibitem [{\citenamefont {Boro{\'n}ska}\ and\ \citenamefont
  {Tuckerman}(2010{\natexlab{a}})}]{boronska2010extreme}%
  \BibitemOpen
  \bibfield  {author} {\bibinfo {author} {\bibfnamefont {K.}~\bibnamefont
  {Boro{\'n}ska}}\ and\ \bibinfo {author} {\bibfnamefont {L.~S.}\ \bibnamefont
  {Tuckerman}},\ }\bibfield  {title} {\bibinfo {title} {{Extreme multiplicity
  in cylindrical Rayleigh-B{\'e}nard convection. I. Time dependence and
  oscillations}},\ }\href {https://link.aps.org/doi/10.1103/PhysRevE.81.036320}
  {\bibfield  {journal} {\bibinfo  {journal} {Phys. Rev. E}\ }\textbf {\bibinfo
  {volume} {81}},\ \bibinfo {pages} {036320} (\bibinfo {year}
  {2010}{\natexlab{a}})}\BibitemShut {NoStop}%
\bibitem [{\citenamefont {Boro{\'n}ska}\ and\ \citenamefont
  {Tuckerman}(2010{\natexlab{b}})}]{boronska2010extreme2}%
  \BibitemOpen
  \bibfield  {author} {\bibinfo {author} {\bibfnamefont {K.}~\bibnamefont
  {Boro{\'n}ska}}\ and\ \bibinfo {author} {\bibfnamefont {L.~S.}\ \bibnamefont
  {Tuckerman}},\ }\bibfield  {title} {\bibinfo {title} {{Extreme multiplicity
  in cylindrical Rayleigh-B{\'e}nard convection. II. Bifurcation diagram and
  symmetry classification}},\ }\href
  {https://link.aps.org/doi/10.1103/PhysRevE.81.036321} {\bibfield  {journal}
  {\bibinfo  {journal} {Phys. Rev. E}\ }\textbf {\bibinfo {volume} {81}},\
  \bibinfo {pages} {036321} (\bibinfo {year} {2010}{\natexlab{b}})}\BibitemShut
  {NoStop}%
\bibitem [{\citenamefont {Arnoldi}(1951)}]{arnoldi1951principle}%
  \BibitemOpen
  \bibfield  {author} {\bibinfo {author} {\bibfnamefont {W.~E.}\ \bibnamefont
  {Arnoldi}},\ }\bibfield  {title} {\bibinfo {title} {The principle of
  minimized iterations in the solution of the matrix eigenvalue problem},\
  }\href {https://www.jstor.org/stable/43633863} {\bibfield  {journal}
  {\bibinfo  {journal} {Q. Appl. Math.}\ }\textbf {\bibinfo {volume} {9}},\
  \bibinfo {pages} {17} (\bibinfo {year} {1951})}\BibitemShut {NoStop}%
\bibitem [{\citenamefont {Puigjaner}\ \emph {et~al.}(2004)\citenamefont
  {Puigjaner}, \citenamefont {Herrero}, \citenamefont {Giralt},\ and\
  \citenamefont {Sim{\'o}}}]{puigjaner2004stability}%
  \BibitemOpen
  \bibfield  {author} {\bibinfo {author} {\bibfnamefont {D.}~\bibnamefont
  {Puigjaner}}, \bibinfo {author} {\bibfnamefont {J.}~\bibnamefont {Herrero}},
  \bibinfo {author} {\bibfnamefont {F.}~\bibnamefont {Giralt}},\ and\ \bibinfo
  {author} {\bibfnamefont {C.}~\bibnamefont {Sim{\'o}}},\ }\bibfield  {title}
  {\bibinfo {title} {Stability analysis of the flow in a cubical cavity heated
  from below},\ }\href {https://doi.org/10.1063/1.1778031} {\bibfield
  {journal} {\bibinfo  {journal} {Phys. Fluids}\ }\textbf {\bibinfo {volume}
  {16}},\ \bibinfo {pages} {3639} (\bibinfo {year} {2004})}\BibitemShut
  {NoStop}%
\bibitem [{\citenamefont {Puigjaner}\ \emph {et~al.}(2006)\citenamefont
  {Puigjaner}, \citenamefont {Herrero}, \citenamefont {Giralt},\ and\
  \citenamefont {Sim{\'o}}}]{puigjaner2006bifurcation}%
  \BibitemOpen
  \bibfield  {author} {\bibinfo {author} {\bibfnamefont {D.}~\bibnamefont
  {Puigjaner}}, \bibinfo {author} {\bibfnamefont {J.}~\bibnamefont {Herrero}},
  \bibinfo {author} {\bibfnamefont {F.}~\bibnamefont {Giralt}},\ and\ \bibinfo
  {author} {\bibfnamefont {C.}~\bibnamefont {Sim{\'o}}},\ }\bibfield  {title}
  {\bibinfo {title} {{Bifurcation analysis of multiple steady flow patterns for
  Rayleigh-B{\'e}nard convection in a cubical cavity at $Pr= 130$}},\ }\href
  {https://link.aps.org/doi/10.1103/PhysRevE.73.046304} {\bibfield  {journal}
  {\bibinfo  {journal} {Phys. Rev. E}\ }\textbf {\bibinfo {volume} {73}},\
  \bibinfo {pages} {046304} (\bibinfo {year} {2006})}\BibitemShut {NoStop}%
\bibitem [{\citenamefont {Doedel}(1981)}]{doedel1981auto}%
  \BibitemOpen
  \bibfield  {author} {\bibinfo {author} {\bibfnamefont {E.~J.}\ \bibnamefont
  {Doedel}},\ }\bibfield  {title} {\bibinfo {title} {{AUTO: A program for the
  automatic bifurcation analysis of autonomous systems}},\ }\href@noop {}
  {\bibfield  {journal} {\bibinfo  {journal} {Congr. Numer}\ }\textbf {\bibinfo
  {volume} {30}},\ \bibinfo {pages} {25} (\bibinfo {year} {1981})}\BibitemShut
  {NoStop}%
\bibitem [{\citenamefont {Uecker}\ \emph {et~al.}(2014)\citenamefont {Uecker},
  \citenamefont {Wetzel},\ and\ \citenamefont
  {Rademacher}}]{uecker2014pde2path}%
  \BibitemOpen
  \bibfield  {author} {\bibinfo {author} {\bibfnamefont {H.}~\bibnamefont
  {Uecker}}, \bibinfo {author} {\bibfnamefont {D.}~\bibnamefont {Wetzel}},\
  and\ \bibinfo {author} {\bibfnamefont {J.~D.~M.}\ \bibnamefont
  {Rademacher}},\ }\bibfield  {title} {\bibinfo {title} {{pde2path-A Matlab
  package for continuation and bifurcation in 2D elliptic systems}},\ }\href
  {https://doi.org/10.1017/S1004897900000295} {\bibfield  {journal} {\bibinfo
  {journal} {Numer. Math.-Theory Me.}\ }\textbf {\bibinfo {volume} {7}},\
  \bibinfo {pages} {58} (\bibinfo {year} {2014})}\BibitemShut {NoStop}%
\bibitem [{\citenamefont {Dijkstra}\ \emph {et~al.}(2014)\citenamefont
  {Dijkstra}, \citenamefont {Wubs}, \citenamefont {Cliffe}, \citenamefont
  {Doedel}, \citenamefont {Dragomirescu}, \citenamefont {Eckhardt},
  \citenamefont {Gelfgat}, \citenamefont {Hazel}, \citenamefont {Lucarini},
  \citenamefont {Salinger} \emph {et~al.}}]{dijkstra2014numerical}%
  \BibitemOpen
  \bibfield  {author} {\bibinfo {author} {\bibfnamefont {H.~A.}\ \bibnamefont
  {Dijkstra}}, \bibinfo {author} {\bibfnamefont {F.~W.}\ \bibnamefont {Wubs}},
  \bibinfo {author} {\bibfnamefont {A.~K.}\ \bibnamefont {Cliffe}}, \bibinfo
  {author} {\bibfnamefont {E.}~\bibnamefont {Doedel}}, \bibinfo {author}
  {\bibfnamefont {I.~F.}\ \bibnamefont {Dragomirescu}}, \bibinfo {author}
  {\bibfnamefont {B.}~\bibnamefont {Eckhardt}}, \bibinfo {author}
  {\bibfnamefont {A.~Y.}\ \bibnamefont {Gelfgat}}, \bibinfo {author}
  {\bibfnamefont {A.~L.}\ \bibnamefont {Hazel}}, \bibinfo {author}
  {\bibfnamefont {V.}~\bibnamefont {Lucarini}}, \bibinfo {author}
  {\bibfnamefont {A.~G.}\ \bibnamefont {Salinger}}, \emph {et~al.},\ }\bibfield
   {title} {\bibinfo {title} {{Numerical Bifurcation Methods and their
  Application to Fluid Dynamics: Analysis beyond Simulation}},\ }\href
  {https://doi.org/10.4208/cicp.240912.180613a} {\bibfield  {journal} {\bibinfo
   {journal} {Commun. Comput. Phys.}\ }\textbf {\bibinfo {volume} {15}},\
  \bibinfo {pages} {1} (\bibinfo {year} {2014})}\BibitemShut {NoStop}%
\bibitem [{\citenamefont {Tuckerman}\ and\ \citenamefont
  {Barkley}(2000)}]{tuckerman2000bifurcation}%
  \BibitemOpen
  \bibfield  {author} {\bibinfo {author} {\bibfnamefont {L.~S.}\ \bibnamefont
  {Tuckerman}}\ and\ \bibinfo {author} {\bibfnamefont {D.}~\bibnamefont
  {Barkley}},\ }\bibfield  {title} {\bibinfo {title} {Bifurcation analysis for
  timesteppers},\ }in\ \href@noop {} {\emph {\bibinfo {booktitle} {Numerical
  methods for bifurcation problems and large-scale dynamical systems}}}\
  (\bibinfo  {publisher} {Springer},\ \bibinfo {year} {2000})\ pp.\ \bibinfo
  {pages} {453--466}\BibitemShut {NoStop}%
\bibitem [{\citenamefont {Mamun}\ and\ \citenamefont
  {Tuckerman}(1995)}]{mamun1995asymmetry}%
  \BibitemOpen
  \bibfield  {author} {\bibinfo {author} {\bibfnamefont {C.~K.}\ \bibnamefont
  {Mamun}}\ and\ \bibinfo {author} {\bibfnamefont {L.~S.}\ \bibnamefont
  {Tuckerman}},\ }\bibfield  {title} {\bibinfo {title} {{Asymmetry and Hopf
  bifurcation in spherical Couette flow}},\ }\href
  {https://doi.org/10.1063/1.868730} {\bibfield  {journal} {\bibinfo  {journal}
  {Phys. Fluids}\ }\textbf {\bibinfo {volume} {7}},\ \bibinfo {pages} {80}
  (\bibinfo {year} {1995})}\BibitemShut {NoStop}%
\bibitem [{\citenamefont {Farrell}\ \emph {et~al.}(2015)\citenamefont
  {Farrell}, \citenamefont {Birkisson},\ and\ \citenamefont
  {Funke}}]{farrell2015deflation}%
  \BibitemOpen
  \bibfield  {author} {\bibinfo {author} {\bibfnamefont {P.~E.}\ \bibnamefont
  {Farrell}}, \bibinfo {author} {\bibfnamefont {A.}~\bibnamefont {Birkisson}},\
  and\ \bibinfo {author} {\bibfnamefont {S.~W.}\ \bibnamefont {Funke}},\
  }\bibfield  {title} {\bibinfo {title} {Deflation techniques for finding
  distinct solutions of nonlinear partial differential equations},\ }\href
  {https://doi.org/10.1137/140984798} {\bibfield  {journal} {\bibinfo
  {journal} {SIAM J. Sci. Comput.}\ }\textbf {\bibinfo {volume} {37}},\
  \bibinfo {pages} {A2026} (\bibinfo {year} {2015})}\BibitemShut {NoStop}%
\bibitem [{\citenamefont {Farrell}\ \emph {et~al.}(2016)\citenamefont
  {Farrell}, \citenamefont {Beentjes},\ and\ \citenamefont
  {Birkisson}}]{farrell2016computation}%
  \BibitemOpen
  \bibfield  {author} {\bibinfo {author} {\bibfnamefont {P.~E.}\ \bibnamefont
  {Farrell}}, \bibinfo {author} {\bibfnamefont {C.~H.}\ \bibnamefont
  {Beentjes}},\ and\ \bibinfo {author} {\bibfnamefont {{\'A}.}~\bibnamefont
  {Birkisson}},\ }\bibfield  {title} {\bibinfo {title} {The computation of
  disconnected bifurcation diagrams},\ }\href
  {https://arxiv.org/abs/1603.00809} {\bibfield  {journal} {\bibinfo  {journal}
  {arXiv preprint arXiv:1603.00809}\ } (\bibinfo {year} {2016})}\BibitemShut
  {NoStop}%
\bibitem [{\citenamefont {Chapman}\ and\ \citenamefont
  {Farrell}(2017)}]{chapman2017analysis}%
  \BibitemOpen
  \bibfield  {author} {\bibinfo {author} {\bibfnamefont {S.~J.}\ \bibnamefont
  {Chapman}}\ and\ \bibinfo {author} {\bibfnamefont {P.~E.}\ \bibnamefont
  {Farrell}},\ }\bibfield  {title} {\bibinfo {title} {{Analysis of Carrier's
  problem}},\ }\href {https://doi.org/10.1137/16M1096074} {\bibfield  {journal}
  {\bibinfo  {journal} {SIAM J. Appl. Math.}\ }\textbf {\bibinfo {volume}
  {77}},\ \bibinfo {pages} {924} (\bibinfo {year} {2017})}\BibitemShut
  {NoStop}%
\bibitem [{\citenamefont {Emerson}\ \emph {et~al.}(2018)\citenamefont
  {Emerson}, \citenamefont {Farrell}, \citenamefont {Adler}, \citenamefont
  {MacLachlan},\ and\ \citenamefont {Atherton}}]{emerson2018computing}%
  \BibitemOpen
  \bibfield  {author} {\bibinfo {author} {\bibfnamefont {D.~B.}\ \bibnamefont
  {Emerson}}, \bibinfo {author} {\bibfnamefont {P.~E.}\ \bibnamefont
  {Farrell}}, \bibinfo {author} {\bibfnamefont {J.~H.}\ \bibnamefont {Adler}},
  \bibinfo {author} {\bibfnamefont {S.~P.}\ \bibnamefont {MacLachlan}},\ and\
  \bibinfo {author} {\bibfnamefont {T.~J.}\ \bibnamefont {Atherton}},\
  }\bibfield  {title} {\bibinfo {title} {Computing equilibrium states of
  cholesteric liquid crystals in elliptical channels with deflation
  algorithms},\ }\href {https://doi.org/10.1080/02678292.2017.1365385}
  {\bibfield  {journal} {\bibinfo  {journal} {Liquid Crystals}\ }\textbf
  {\bibinfo {volume} {45}},\ \bibinfo {pages} {341} (\bibinfo {year}
  {2018})}\BibitemShut {NoStop}%
\bibitem [{\citenamefont {Charalampidis}\ \emph {et~al.}(2018)\citenamefont
  {Charalampidis}, \citenamefont {Kevrekidis},\ and\ \citenamefont
  {Farrell}}]{charalampidis2018computing}%
  \BibitemOpen
  \bibfield  {author} {\bibinfo {author} {\bibfnamefont {E.~G.}\ \bibnamefont
  {Charalampidis}}, \bibinfo {author} {\bibfnamefont {P.~G.}\ \bibnamefont
  {Kevrekidis}},\ and\ \bibinfo {author} {\bibfnamefont {P.~E.}\ \bibnamefont
  {Farrell}},\ }\bibfield  {title} {\bibinfo {title} {Computing stationary
  solutions of the two-dimensional {G}ross--{P}itaevskii equation with deflated
  continuation},\ }\href
  {https://doi.org/https://doi.org/10.1016/j.cnsns.2017.05.024} {\bibfield
  {journal} {\bibinfo  {journal} {Commun. Nonlinear Sci. Numer. Simulat.}\
  }\textbf {\bibinfo {volume} {54}},\ \bibinfo {pages} {482} (\bibinfo {year}
  {2018})}\BibitemShut {NoStop}%
\bibitem [{\citenamefont {Charalampidis}\ \emph {et~al.}(2020)\citenamefont
  {Charalampidis}, \citenamefont {Boull\'e}, \citenamefont {Kevrekidis},\ and\
  \citenamefont {Farrell}}]{charalampidis2019bifurcation}%
  \BibitemOpen
  \bibfield  {author} {\bibinfo {author} {\bibfnamefont {E.~G.}\ \bibnamefont
  {Charalampidis}}, \bibinfo {author} {\bibfnamefont {N.}~\bibnamefont
  {Boull\'e}}, \bibinfo {author} {\bibfnamefont {P.~G.}\ \bibnamefont
  {Kevrekidis}},\ and\ \bibinfo {author} {\bibfnamefont {P.~E.}\ \bibnamefont
  {Farrell}},\ }\bibfield  {title} {\bibinfo {title} {Bifurcation analysis of
  stationary solutions of two-dimensional coupled {G}ross--{P}itaevskii
  equations using deflated continuation},\ }\href
  {https://doi.org/10.1016/j.cnsns.2020.105255} {\bibfield  {journal} {\bibinfo
   {journal} {Commun. Nonlinear Sci. Numer. Simulat.}\ }\textbf {\bibinfo
  {volume} {87}},\ \bibinfo {pages} {105255} (\bibinfo {year}
  {2020})}\BibitemShut {NoStop}%
\bibitem [{\citenamefont {Boull{\'e}}\ \emph {et~al.}(2020)\citenamefont
  {Boull{\'e}}, \citenamefont {Charalampidis}, \citenamefont {Farrell},\ and\
  \citenamefont {Kevrekidis}}]{boulle2020deflation}%
  \BibitemOpen
  \bibfield  {author} {\bibinfo {author} {\bibfnamefont {N.}~\bibnamefont
  {Boull{\'e}}}, \bibinfo {author} {\bibfnamefont {E.~G.}\ \bibnamefont
  {Charalampidis}}, \bibinfo {author} {\bibfnamefont {P.~E.}\ \bibnamefont
  {Farrell}},\ and\ \bibinfo {author} {\bibfnamefont {P.~G.}\ \bibnamefont
  {Kevrekidis}},\ }\bibfield  {title} {\bibinfo {title} {{Deflation-based
  identification of nonlinear excitations of the three-dimensional
  Gross-Pitaevskii equation}},\ }\href
  {https://link.aps.org/doi/10.1103/PhysRevA.102.053307} {\bibfield  {journal}
  {\bibinfo  {journal} {Phys. Rev. A}\ }\textbf {\bibinfo {volume} {102}},\
  \bibinfo {pages} {053307} (\bibinfo {year} {2020})}\BibitemShut {NoStop}%
\bibitem [{\citenamefont {Fauve}\ \emph {et~al.}(2017)\citenamefont {Fauve},
  \citenamefont {Herault}, \citenamefont {Michel},\ and\ \citenamefont
  {P{\'e}tr{\'e}lis}}]{fauveetal17}%
  \BibitemOpen
  \bibfield  {author} {\bibinfo {author} {\bibfnamefont {S.}~\bibnamefont
  {Fauve}}, \bibinfo {author} {\bibfnamefont {J.}~\bibnamefont {Herault}},
  \bibinfo {author} {\bibfnamefont {G.}~\bibnamefont {Michel}},\ and\ \bibinfo
  {author} {\bibfnamefont {F.}~\bibnamefont {P{\'e}tr{\'e}lis}},\ }\bibfield
  {title} {\bibinfo {title} {Instabilities on a turbulent background},\ }\href
  {https://doi.org/10.1088/1742-5468/aa6f3d} {\bibfield  {journal} {\bibinfo
  {journal} {J. Stat. Mech. Theory Exp.}\ }\textbf {\bibinfo {volume} {2017}},\
  \bibinfo {pages} {064001} (\bibinfo {year} {2017})}\BibitemShut {NoStop}%
\bibitem [{\citenamefont {Kadanoff}(2000)}]{kadanoff00}%
  \BibitemOpen
  \bibfield  {author} {\bibinfo {author} {\bibfnamefont {L.~P.}\ \bibnamefont
  {Kadanoff}},\ }\href@noop {} {\emph {\bibinfo {title} {Statistical physics:
  statics, dynamics and renormalization}}}\ (\bibinfo  {publisher} {World
  Scientific Publishing Company},\ \bibinfo {year} {2000})\BibitemShut
  {NoStop}%
\bibitem [{\citenamefont {Sugiyama}\ \emph {et~al.}(2010)\citenamefont
  {Sugiyama}, \citenamefont {Ni}, \citenamefont {Stevens}, \citenamefont
  {Chan}, \citenamefont {Zhou}, \citenamefont {Xi}, \citenamefont {Sun},
  \citenamefont {Grossmann}, \citenamefont {Xia},\ and\ \citenamefont
  {Lohse}}]{sugiyamaetal10}%
  \BibitemOpen
  \bibfield  {author} {\bibinfo {author} {\bibfnamefont {K.}~\bibnamefont
  {Sugiyama}}, \bibinfo {author} {\bibfnamefont {R.}~\bibnamefont {Ni}},
  \bibinfo {author} {\bibfnamefont {R.~J.}\ \bibnamefont {Stevens}}, \bibinfo
  {author} {\bibfnamefont {T.~S.}\ \bibnamefont {Chan}}, \bibinfo {author}
  {\bibfnamefont {S.-Q.}\ \bibnamefont {Zhou}}, \bibinfo {author}
  {\bibfnamefont {H.-D.}\ \bibnamefont {Xi}}, \bibinfo {author} {\bibfnamefont
  {C.}~\bibnamefont {Sun}}, \bibinfo {author} {\bibfnamefont {S.}~\bibnamefont
  {Grossmann}}, \bibinfo {author} {\bibfnamefont {K.-Q.}\ \bibnamefont {Xia}},\
  and\ \bibinfo {author} {\bibfnamefont {D.}~\bibnamefont {Lohse}},\ }\bibfield
   {title} {\bibinfo {title} {Flow reversals in thermally driven turbulence},\
  }\href {https://link.aps.org/doi/10.1103/PhysRevLett.105.034503} {\bibfield
  {journal} {\bibinfo  {journal} {Phys. Rev. Lett.}\ }\textbf {\bibinfo
  {volume} {105}},\ \bibinfo {pages} {034503} (\bibinfo {year}
  {2010})}\BibitemShut {NoStop}%
\bibitem [{\citenamefont {Chandra}\ and\ \citenamefont
  {Verma}(2013)}]{chandraverma13}%
  \BibitemOpen
  \bibfield  {author} {\bibinfo {author} {\bibfnamefont {M.}~\bibnamefont
  {Chandra}}\ and\ \bibinfo {author} {\bibfnamefont {M.~K.}\ \bibnamefont
  {Verma}},\ }\bibfield  {title} {\bibinfo {title} {Flow reversals in turbulent
  convection via vortex reconnections},\ }\href
  {https://link.aps.org/doi/10.1103/PhysRevLett.110.114503} {\bibfield
  {journal} {\bibinfo  {journal} {Phys. Rev. Lett.}\ }\textbf {\bibinfo
  {volume} {110}},\ \bibinfo {pages} {114503} (\bibinfo {year}
  {2013})}\BibitemShut {NoStop}%
\bibitem [{\citenamefont {Podvin}\ and\ \citenamefont
  {Sergent}(2017)}]{podvinsergent17}%
  \BibitemOpen
  \bibfield  {author} {\bibinfo {author} {\bibfnamefont {B.}~\bibnamefont
  {Podvin}}\ and\ \bibinfo {author} {\bibfnamefont {A.}~\bibnamefont
  {Sergent}},\ }\bibfield  {title} {\bibinfo {title} {{Precursor for wind
  reversal in a square Rayleigh-B{\'e}nard cell}},\ }\href
  {https://link.aps.org/doi/10.1103/PhysRevE.95.013112} {\bibfield  {journal}
  {\bibinfo  {journal} {Phys. Rev. E}\ }\textbf {\bibinfo {volume} {95}},\
  \bibinfo {pages} {013112} (\bibinfo {year} {2017})}\BibitemShut {NoStop}%
\bibitem [{\citenamefont {B{\'e}nard}(1900)}]{benard1900}%
  \BibitemOpen
  \bibfield  {author} {\bibinfo {author} {\bibfnamefont {H.}~\bibnamefont
  {B{\'e}nard}},\ }\bibfield  {title} {\bibinfo {title} {{Etude
  exp{\'e}rimentale du mouvement des liquides propageant de la chaleur par
  convection. R{\'e}gime permanent: tourbillons cellulaires}},\ }\href@noop {}
  {\bibfield  {journal} {\bibinfo  {journal} {C. r. hebd. s\'eances Acad. sci.
  Paris}\ }\textbf {\bibinfo {volume} {130}},\ \bibinfo {pages} {1004}
  (\bibinfo {year} {1900})}\BibitemShut {NoStop}%
\bibitem [{\citenamefont {Rayleigh}(1916)}]{rayleigh1916}%
  \BibitemOpen
  \bibfield  {author} {\bibinfo {author} {\bibfnamefont {L.}~\bibnamefont
  {Rayleigh}},\ }\bibfield  {title} {\bibinfo {title} {{On convection currents
  in a horizontal layer of fluid, when the higher temperature is on the under
  side}},\ }\href {https://doi.org/10.1080/14786441608635602} {\bibfield
  {journal} {\bibinfo  {journal} {Phil. Mag. S.}\ }\textbf {\bibinfo {volume}
  {32}},\ \bibinfo {pages} {529} (\bibinfo {year} {1916})}\BibitemShut
  {NoStop}%
\bibitem [{\citenamefont {B{\'e}nard}(1927)}]{benard1927}%
  \BibitemOpen
  \bibfield  {author} {\bibinfo {author} {\bibfnamefont {H.}~\bibnamefont
  {B{\'e}nard}},\ }\bibfield  {title} {\bibinfo {title} {{Sur les tourbillons
  cellulaires et la th{\'e}orie de Rayleigh}},\ }\href@noop {} {\bibfield
  {journal} {\bibinfo  {journal} {C. r. hebd. s\'eances Acad. sci. Paris}\
  }\textbf {\bibinfo {volume} {185}},\ \bibinfo {pages} {1109} (\bibinfo {year}
  {1927})}\BibitemShut {NoStop}%
\bibitem [{\citenamefont {Oberbeck}(1879)}]{oberbeck1879}%
  \BibitemOpen
  \bibfield  {author} {\bibinfo {author} {\bibfnamefont {A.}~\bibnamefont
  {Oberbeck}},\ }\bibfield  {title} {\bibinfo {title} {{{\"U}ber die
  W{\"a}rmeleitung der Fl{\"u}ssigkeiten bei Ber{\"u}cksichtigung der
  Str{\"o}mungen infolge von Temperaturdifferenzen}},\ }\href
  {https://doi.org/10.1002/andp.18792430606} {\bibfield  {journal} {\bibinfo
  {journal} {Ann. Phys. Chem.}\ }\textbf {\bibinfo {volume} {243}},\ \bibinfo
  {pages} {271} (\bibinfo {year} {1879})}\BibitemShut {NoStop}%
\bibitem [{\citenamefont {Boussinesq}(1903)}]{boussinesq1903}%
  \BibitemOpen
  \bibfield  {author} {\bibinfo {author} {\bibfnamefont {J.}~\bibnamefont
  {Boussinesq}},\ }\href@noop {} {\emph {\bibinfo {title} {{Th{\'e}orie
  analytique de la chaleur mise en harmonic avec la thermodynamique et avec la
  th{\'e}orie m{\'e}canique de la lumi{\`e}re}}}},\ Vol.~\bibinfo {volume}
  {II}\ (\bibinfo  {publisher} {Gauthier-Villars},\ \bibinfo {year}
  {1903})\BibitemShut {NoStop}%
\bibitem [{\citenamefont {Tritton}(2012)}]{tritton12}%
  \BibitemOpen
  \bibfield  {author} {\bibinfo {author} {\bibfnamefont {D.~J.}\ \bibnamefont
  {Tritton}},\ }\href@noop {} {\emph {\bibinfo {title} {Physical fluid
  dynamics}}}\ (\bibinfo  {publisher} {Springer},\ \bibinfo {year}
  {2012})\BibitemShut {NoStop}%
\bibitem [{\citenamefont {Rathgeber}\ \emph {et~al.}(2016)\citenamefont
  {Rathgeber}, \citenamefont {Ham}, \citenamefont {Mitchell}, \citenamefont
  {Lange}, \citenamefont {Luporini}, \citenamefont {Mcrae}, \citenamefont
  {Bercea}, \citenamefont {Markall},\ and\ \citenamefont
  {Kelly}}]{rathgeber2016}%
  \BibitemOpen
  \bibfield  {author} {\bibinfo {author} {\bibfnamefont {F.}~\bibnamefont
  {Rathgeber}}, \bibinfo {author} {\bibfnamefont {D.~A.}\ \bibnamefont {Ham}},
  \bibinfo {author} {\bibfnamefont {L.}~\bibnamefont {Mitchell}}, \bibinfo
  {author} {\bibfnamefont {M.}~\bibnamefont {Lange}}, \bibinfo {author}
  {\bibfnamefont {F.}~\bibnamefont {Luporini}}, \bibinfo {author}
  {\bibfnamefont {A.~T.~T.}\ \bibnamefont {Mcrae}}, \bibinfo {author}
  {\bibfnamefont {G.-T.}\ \bibnamefont {Bercea}}, \bibinfo {author}
  {\bibfnamefont {G.~R.}\ \bibnamefont {Markall}},\ and\ \bibinfo {author}
  {\bibfnamefont {P.~H.~J.}\ \bibnamefont {Kelly}},\ }\bibfield  {title}
  {\bibinfo {title} {Firedrake: automating the finite element method by
  composing abstractions},\ }\href {https://doi.org/10.1145/2998441} {\bibfield
   {journal} {\bibinfo  {journal} {ACM Trans. Math. Softw.}\ }\textbf {\bibinfo
  {volume} {43}},\ \bibinfo {pages} {1} (\bibinfo {year} {2016})}\BibitemShut
  {NoStop}%
\bibitem [{\citenamefont {Amestoy}\ \emph {et~al.}(2000)\citenamefont
  {Amestoy}, \citenamefont {Duff}, \citenamefont {L'Excellent},\ and\
  \citenamefont {Koster}}]{amestoy2000mumps}%
  \BibitemOpen
  \bibfield  {author} {\bibinfo {author} {\bibfnamefont {P.~R.}\ \bibnamefont
  {Amestoy}}, \bibinfo {author} {\bibfnamefont {I.~S.}\ \bibnamefont {Duff}},
  \bibinfo {author} {\bibfnamefont {J.-Y.}\ \bibnamefont {L'Excellent}},\ and\
  \bibinfo {author} {\bibfnamefont {J.}~\bibnamefont {Koster}},\ }\bibfield
  {title} {\bibinfo {title} {{MUMPS: a general purpose distributed memory
  sparse solver}},\ }in\ \href@noop {} {\emph {\bibinfo {booktitle}
  {International Workshop on Applied Parallel Computing}}}\ (\bibinfo
  {organization} {Springer},\ \bibinfo {year} {2000})\ pp.\ \bibinfo {pages}
  {121--130}\BibitemShut {NoStop}%
\bibitem [{\citenamefont {Stewart}(2002)}]{stewart2002}%
  \BibitemOpen
  \bibfield  {author} {\bibinfo {author} {\bibfnamefont {G.~W.}\ \bibnamefont
  {Stewart}},\ }\bibfield  {title} {\bibinfo {title} {A {K}rylov--{S}chur
  algorithm for large eigenproblems},\ }\href
  {https://doi.org/10.1137/S0895479800371529} {\bibfield  {journal} {\bibinfo
  {journal} {SIAM J. Matrix Anal. A.}\ }\textbf {\bibinfo {volume} {23}},\
  \bibinfo {pages} {601} (\bibinfo {year} {2002})}\BibitemShut {NoStop}%
\bibitem [{\citenamefont {Hernandez}\ \emph {et~al.}(2005)\citenamefont
  {Hernandez}, \citenamefont {Roman},\ and\ \citenamefont
  {Vidal}}]{hernandez2005slepc}%
  \BibitemOpen
  \bibfield  {author} {\bibinfo {author} {\bibfnamefont {V.}~\bibnamefont
  {Hernandez}}, \bibinfo {author} {\bibfnamefont {J.~E.}\ \bibnamefont
  {Roman}},\ and\ \bibinfo {author} {\bibfnamefont {V.}~\bibnamefont {Vidal}},\
  }\bibfield  {title} {\bibinfo {title} {{SLEP}c: {A} scalable and flexible
  toolkit for the solution of eigenvalue problems},\ }\href
  {https://doi.org/10.1145/1089014.1089019} {\bibfield  {journal} {\bibinfo
  {journal} {ACM Trans. Math. Softw.}\ }\textbf {\bibinfo {volume} {31}},\
  \bibinfo {pages} {351} (\bibinfo {year} {2005})}\BibitemShut {NoStop}%
\bibitem [{\citenamefont
  {Chandrasekhar}(1961)}]{chandrasekhar1961hydrodynamic}%
  \BibitemOpen
  \bibfield  {author} {\bibinfo {author} {\bibfnamefont {S.}~\bibnamefont
  {Chandrasekhar}},\ }\href@noop {} {\emph {\bibinfo {title} {{Hydrodynamic and
  Hydromagnetic Stability}}}}\ (\bibinfo  {publisher} {Oxford University
  Press},\ \bibinfo {year} {1961})\BibitemShut {NoStop}%
\bibitem [{\citenamefont {Fauve}(2017)}]{fauve17}%
  \BibitemOpen
  \bibfield  {author} {\bibinfo {author} {\bibfnamefont {S.}~\bibnamefont
  {Fauve}},\ }\bibfield  {title} {\bibinfo {title} {{Henri B{\'e}nard and
  pattern-forming instabilities}},\ }\href
  {https://doi.org/10.1016/j.crhy.2017.11.002} {\bibfield  {journal} {\bibinfo
  {journal} {C. R. Phys.}\ }\textbf {\bibinfo {volume} {18}},\ \bibinfo {pages}
  {531} (\bibinfo {year} {2017})}\BibitemShut {NoStop}%
\bibitem [{\citenamefont {Mizushima}(1995)}]{mizushima95}%
  \BibitemOpen
  \bibfield  {author} {\bibinfo {author} {\bibfnamefont {J.}~\bibnamefont
  {Mizushima}},\ }\bibfield  {title} {\bibinfo {title} {{Onset of the Thermal
  Convection in a Finite Two-Dimensional Box}},\ }\href
  {https://doi.org/10.1143/JPSJ.64.2420} {\bibfield  {journal} {\bibinfo
  {journal} {J. Phys. Soc. Japan}\ }\textbf {\bibinfo {volume} {64}},\ \bibinfo
  {pages} {2420} (\bibinfo {year} {1995})}\BibitemShut {NoStop}%
\bibitem [{\citenamefont {Siggia}(1994)}]{siggia94}%
  \BibitemOpen
  \bibfield  {author} {\bibinfo {author} {\bibfnamefont {E.~D.}\ \bibnamefont
  {Siggia}},\ }\bibfield  {title} {\bibinfo {title} {{High Rayleigh number
  convection}},\ }\href {https://doi.org/10.1146/annurev.fl.26.010194.001033}
  {\bibfield  {journal} {\bibinfo  {journal} {Annu. Rev. Fluid Mech.}\ }\textbf
  {\bibinfo {volume} {26}},\ \bibinfo {pages} {137} (\bibinfo {year}
  {1994})}\BibitemShut {NoStop}%
\bibitem [{\citenamefont {Kuznetsov}(2013)}]{kuznetsov13}%
  \BibitemOpen
  \bibfield  {author} {\bibinfo {author} {\bibfnamefont {Y.~A.}\ \bibnamefont
  {Kuznetsov}},\ }\href@noop {} {\emph {\bibinfo {title} {Elements of applied
  bifurcation theory}}},\ Vol.\ \bibinfo {volume} {112}\ (\bibinfo  {publisher}
  {Springer Science \& Business Media},\ \bibinfo {year} {2013})\BibitemShut
  {NoStop}%
\bibitem [{\citenamefont {Kadanoff}(2001)}]{kadanoff01}%
  \BibitemOpen
  \bibfield  {author} {\bibinfo {author} {\bibfnamefont {L.~P.}\ \bibnamefont
  {Kadanoff}},\ }\bibfield  {title} {\bibinfo {title} {Turbulent heat flow:
  structures and scaling},\ }\href {https://doi.org/10.1063/1.1404847}
  {\bibfield  {journal} {\bibinfo  {journal} {Phys. Today}\ }\textbf {\bibinfo
  {volume} {54}},\ \bibinfo {pages} {34} (\bibinfo {year} {2001})}\BibitemShut
  {NoStop}%
\bibitem [{\citenamefont {Ravelet}\ \emph {et~al.}(2004)\citenamefont
  {Ravelet}, \citenamefont {Mari{\'e}}, \citenamefont {Chiffaudel},\ and\
  \citenamefont {Daviaud}}]{raveletetal04}%
  \BibitemOpen
  \bibfield  {author} {\bibinfo {author} {\bibfnamefont {F.}~\bibnamefont
  {Ravelet}}, \bibinfo {author} {\bibfnamefont {L.}~\bibnamefont {Mari{\'e}}},
  \bibinfo {author} {\bibfnamefont {A.}~\bibnamefont {Chiffaudel}},\ and\
  \bibinfo {author} {\bibfnamefont {F.}~\bibnamefont {Daviaud}},\ }\bibfield
  {title} {\bibinfo {title} {{Multistability and memory effect in a highly
  turbulent flow: Experimental evidence for a global bifurcation}},\ }\href
  {https://link.aps.org/doi/10.1103/PhysRevLett.93.164501} {\bibfield
  {journal} {\bibinfo  {journal} {Phys. Rev. Lett.}\ }\textbf {\bibinfo
  {volume} {93}},\ \bibinfo {pages} {164501} (\bibinfo {year}
  {2004})}\BibitemShut {NoStop}%
\bibitem [{\citenamefont {Berhanu}\ \emph {et~al.}(2007)\citenamefont
  {Berhanu}, \citenamefont {Monchaux}, \citenamefont {Fauve}, \citenamefont
  {Mordant}, \citenamefont {P{\'e}tr{\'e}lis}, \citenamefont {Chiffaudel},
  \citenamefont {Daviaud}, \citenamefont {Dubrulle}, \citenamefont {Mari{\'e}},
  \citenamefont {Ravelet} \emph {et~al.}}]{berhanuetal07}%
  \BibitemOpen
  \bibfield  {author} {\bibinfo {author} {\bibfnamefont {M.}~\bibnamefont
  {Berhanu}}, \bibinfo {author} {\bibfnamefont {R.}~\bibnamefont {Monchaux}},
  \bibinfo {author} {\bibfnamefont {S.}~\bibnamefont {Fauve}}, \bibinfo
  {author} {\bibfnamefont {N.}~\bibnamefont {Mordant}}, \bibinfo {author}
  {\bibfnamefont {F.}~\bibnamefont {P{\'e}tr{\'e}lis}}, \bibinfo {author}
  {\bibfnamefont {A.}~\bibnamefont {Chiffaudel}}, \bibinfo {author}
  {\bibfnamefont {F.}~\bibnamefont {Daviaud}}, \bibinfo {author} {\bibfnamefont
  {B.}~\bibnamefont {Dubrulle}}, \bibinfo {author} {\bibfnamefont
  {L.}~\bibnamefont {Mari{\'e}}}, \bibinfo {author} {\bibfnamefont
  {F.}~\bibnamefont {Ravelet}}, \emph {et~al.},\ }\bibfield  {title} {\bibinfo
  {title} {Magnetic field reversals in an experimental turbulent dynamo},\
  }\href {https://doi.org/10.1209/0295-5075/77/59001} {\bibfield  {journal}
  {\bibinfo  {journal} {EPL-Europhys. Lett.}\ }\textbf {\bibinfo {volume}
  {77}},\ \bibinfo {pages} {59001} (\bibinfo {year} {2007})}\BibitemShut
  {NoStop}%
\bibitem [{\citenamefont {Cadot}\ \emph {et~al.}(2015)\citenamefont {Cadot},
  \citenamefont {Evrard},\ and\ \citenamefont {Pastur}}]{cadotetal15}%
  \BibitemOpen
  \bibfield  {author} {\bibinfo {author} {\bibfnamefont {O.}~\bibnamefont
  {Cadot}}, \bibinfo {author} {\bibfnamefont {A.}~\bibnamefont {Evrard}},\ and\
  \bibinfo {author} {\bibfnamefont {L.}~\bibnamefont {Pastur}},\ }\bibfield
  {title} {\bibinfo {title} {Imperfect supercritical bifurcation in a
  three-dimensional turbulent wake},\ }\href
  {https://link.aps.org/doi/10.1103/PhysRevE.91.063005} {\bibfield  {journal}
  {\bibinfo  {journal} {Phys. Rev. E}\ }\textbf {\bibinfo {volume} {91}},\
  \bibinfo {pages} {063005} (\bibinfo {year} {2015})}\BibitemShut {NoStop}%
\bibitem [{\citenamefont {Dallas}\ \emph {et~al.}(2020)\citenamefont {Dallas},
  \citenamefont {Seshasayanan},\ and\ \citenamefont {Fauve}}]{dallasetal19}%
  \BibitemOpen
  \bibfield  {author} {\bibinfo {author} {\bibfnamefont {V.}~\bibnamefont
  {Dallas}}, \bibinfo {author} {\bibfnamefont {K.}~\bibnamefont
  {Seshasayanan}},\ and\ \bibinfo {author} {\bibfnamefont {S.}~\bibnamefont
  {Fauve}},\ }\bibfield  {title} {\bibinfo {title} {Transitions between
  turbulent states in a two-dimensional shear flow},\ }\href
  {https://link.aps.org/doi/10.1103/PhysRevFluids.5.084610} {\bibfield
  {journal} {\bibinfo  {journal} {Phys. Rev. Fluids}\ }\textbf {\bibinfo
  {volume} {5}},\ \bibinfo {pages} {084610} (\bibinfo {year}
  {2020})}\BibitemShut {NoStop}%
\bibitem [{\citenamefont {Winchester}\ \emph {et~al.}(2021)\citenamefont
  {Winchester}, \citenamefont {Dallas},\ and\ \citenamefont {Howell}}]{wdp20}%
  \BibitemOpen
  \bibfield  {author} {\bibinfo {author} {\bibfnamefont {P.}~\bibnamefont
  {Winchester}}, \bibinfo {author} {\bibfnamefont {V.}~\bibnamefont {Dallas}},\
  and\ \bibinfo {author} {\bibfnamefont {P.~D.}\ \bibnamefont {Howell}},\
  }\bibfield  {title} {\bibinfo {title} {{Zonal flow reversals in
  two-dimensional Rayleigh-B\'enard convection}},\ }\href
  {https://doi.org/10.1103/PhysRevFluids.6.033502} {\bibfield  {journal}
  {\bibinfo  {journal} {Phys. Rev. Fluids}\ }\textbf {\bibinfo {volume} {6}},\
  \bibinfo {pages} {033502} (\bibinfo {year} {2021})}\BibitemShut {NoStop}%
\bibitem [{\citenamefont {Sondak}\ \emph {et~al.}(2015)\citenamefont {Sondak},
  \citenamefont {Smith},\ and\ \citenamefont {Waleffe}}]{sondaketak15}%
  \BibitemOpen
  \bibfield  {author} {\bibinfo {author} {\bibfnamefont {D.}~\bibnamefont
  {Sondak}}, \bibinfo {author} {\bibfnamefont {L.~M.}\ \bibnamefont {Smith}},\
  and\ \bibinfo {author} {\bibfnamefont {F.}~\bibnamefont {Waleffe}},\
  }\bibfield  {title} {\bibinfo {title} {{Optimal heat transport solutions for
  Rayleigh--B{\'e}nard convection}},\ }\href
  {https://doi.org/10.1017/jfm.2015.615} {\bibfield  {journal} {\bibinfo
  {journal} {J. Fluid Mech.}\ }\textbf {\bibinfo {volume} {784}},\ \bibinfo
  {pages} {565} (\bibinfo {year} {2015})}\BibitemShut {NoStop}%
\bibitem [{\citenamefont {Waleffe}\ \emph {et~al.}(2015)\citenamefont
  {Waleffe}, \citenamefont {Boonkasame},\ and\ \citenamefont
  {Smith}}]{waleffeetal15}%
  \BibitemOpen
  \bibfield  {author} {\bibinfo {author} {\bibfnamefont {F.}~\bibnamefont
  {Waleffe}}, \bibinfo {author} {\bibfnamefont {A.}~\bibnamefont
  {Boonkasame}},\ and\ \bibinfo {author} {\bibfnamefont {L.~M.}\ \bibnamefont
  {Smith}},\ }\bibfield  {title} {\bibinfo {title} {{Heat transport by coherent
  Rayleigh-B{\'e}nard convection}},\ }\href {https://doi.org/10.1063/1.4919930}
  {\bibfield  {journal} {\bibinfo  {journal} {Phys. Fluids}\ }\textbf {\bibinfo
  {volume} {27}},\ \bibinfo {pages} {051702} (\bibinfo {year}
  {2015})}\BibitemShut {NoStop}%
\bibitem [{\citenamefont {Farrell}\ and\ \citenamefont
  {Gazca-Orozco}(2020)}]{farrell2020augmented}%
  \BibitemOpen
  \bibfield  {author} {\bibinfo {author} {\bibfnamefont {P.~E.}\ \bibnamefont
  {Farrell}}\ and\ \bibinfo {author} {\bibfnamefont {P.~A.}\ \bibnamefont
  {Gazca-Orozco}},\ }\bibfield  {title} {\bibinfo {title} {{An augmented
  Lagrangian preconditioner for implicitly-constituted non-Newtonian
  incompressible flow}},\ }\href {https://doi.org/10.1137/20M1336618}
  {\bibfield  {journal} {\bibinfo  {journal} {SIAM J. Sci. Comput.}\ }\textbf
  {\bibinfo {volume} {42}},\ \bibinfo {pages} {B1329} (\bibinfo {year}
  {2020})}\BibitemShut {NoStop}%
\bibitem [{\citenamefont {Farrell}\ \emph {et~al.}(2019)\citenamefont
  {Farrell}, \citenamefont {Mitchell},\ and\ \citenamefont
  {Wechsung}}]{farrell2019augmented}%
  \BibitemOpen
  \bibfield  {author} {\bibinfo {author} {\bibfnamefont {P.~E.}\ \bibnamefont
  {Farrell}}, \bibinfo {author} {\bibfnamefont {L.}~\bibnamefont {Mitchell}},\
  and\ \bibinfo {author} {\bibfnamefont {F.}~\bibnamefont {Wechsung}},\
  }\bibfield  {title} {\bibinfo {title} {{An Augmented Lagrangian
  Preconditioner for the 3D Stationary Incompressible Navier--Stokes Equations
  at High Reynolds Number}},\ }\href {https://doi.org/10.1137/18M1219370}
  {\bibfield  {journal} {\bibinfo  {journal} {SIAM J. Sci. Comput.}\ }\textbf
  {\bibinfo {volume} {41}},\ \bibinfo {pages} {A3073} (\bibinfo {year}
  {2019})}\BibitemShut {NoStop}%
\end{thebibliography}%

\end{document}